\documentclass[final,12pt]{elsarticle}

\usepackage{amssymb}
\usepackage{amsmath}
\usepackage[linesnumbered,ruled,vlined]{algorithm2e}

\usepackage{multirow}
\usepackage{url}
\usepackage{float}
\usepackage{epsfig}
\usepackage{amsfonts}
\usepackage{balance}
\usepackage{booktabs}
\usepackage{amsmath}
\usepackage{graphicx}
\usepackage{pdfpages}
\usepackage{subfigure}
\usepackage{epstopdf}
\usepackage{tabularx}
\usepackage{changepage}

\journal{Information Sciences}

\begin{document}

\begin{frontmatter}
\title{GLIMG: Global and Local Item Graphs for Top-N Recommender Systems}


\author[]{Zhuoyi Lin\fnref{label1}}
\ead{ZHUOYI001@ntu.edu.sg}

\author{Lei Feng\corref{mycorrespondingauthor}\fnref{label2,label3}}
\cortext[mycorrespondingauthor]{Corresponding author}
\ead{lfeng@cqu.edu.cn}

\author{Rui Yin\fnref{label1}}
\ead{YINR0002@e.ntu.edu.sg}
\author{Chi Xu\fnref{label1,label4}}
\ead{cxu@simtech.a-star.edu.sg}

\author{Chee Keong Kwoh\fnref{label1}}
\ead{ASCKKWOH@ntu.edu.sg}

\address[label1]{School of Computer Science and Engineering, Nanyang Technological University, Singapore 639798}
\address[label2]{College of Computer Science, Chongqing University, China}
\address[label3]{RIKEN Center for Advanced Intelligence Project, Japan}
\address[label4]{Singapore Institute of Manufacturing Technology, Singapore 138634}







\begin{abstract}
Graph-based recommendation models work well for top-N recommender systems due to their capability to capture the potential relationships between entities. However, most of the existing methods only construct a single global item graph shared by all the users and regrettably ignore the diverse tastes between different user groups. Inspired by the success of local models for recommendation, this paper provides the first attempt to investigate multiple local item graphs along with a global item graph for graph-based recommendation models. We argue that recommendation on global and local graphs outperforms that on a single global graph or multiple local graphs. Specifically, we propose a novel graph-based recommendation model named GLIMG (\underline{G}lobal and \underline{L}ocal \underline{I}te\underline{M} \underline{G}raphs), which simultaneously captures both the global and local user tastes. By integrating the global and local graphs into an adapted semi-supervised learning model, users' preferences on items are propagated globally and locally. Extensive experimental results on real-world datasets show that our proposed method consistently outperforms the state-of-the-art counterparts on the top-N recommendation task.
\end{abstract}




\end{frontmatter}


\section{INTRODUCTION}

Due to the prosperity and development of e-commerce, recommender systems now play an increasingly significant role. For companies (e.g., Amazon, Netflix and Alibaba), recommender systems are widely used to target potential customers. In reality, top-N recommender systems are very popular, which aim to generate a meaningful list of items for users \cite{performance2010, LCR}.

Among the existing recommendation techniques, collaborative filtering (CF) methods are widely adopted and achieve impressive performance. CF-based methods can be generally categorized into two groups: latent factor models and neighborhood methods \cite{FMN}. Latent factor models work by modeling users and items using latent factors while neighborhood methods aim to estimate the relationships between users (i.e. user-based methods) or items (i.e. item-based methods).

As a neighborhood-based method, the graph-based recommendation approach has attracted increasing attention in recommender systems. It alleviates the data sparsity problem by capturing the structural transitivity when defining the similarity among users or items\cite{HF, birank}.
In graph-based methods, users or items are usually denoted as nodes, the relationships between users or items are represented by edges and the preferences of users for items are propagated on the nodes.
Since item-based methods have been shown to achieve satisfied performance and exhibit high scalability to the top-N recommendation task compared with user-based methods \cite{glslim,fism,slim,itemcf}, graph-based models normally rely on the item graph \cite{itemrank,itemgraph}. However, most of the existing graph-based models only construct a single global item graph shared by all the users and regrettably ignore the diverse tastes between different user groups.

Inspired by the development of local models in the recommender systems, in this paper, we propose to solve the above problem by constructing multiple local item graphs along with a global item graphs for the graph-based model. Our method GLIMG (\underline{G}lobal and \underline{L}ocal \underline{I}te\underline{M} \underline{G}raphs) captures not only the global taste shared by all the users but also the local tastes of different user subgroups. Extensive experiments show that our model outperforms competing top-N recommendation models on real world datasets.

The main contributions of this study are summarized as follows:

\begin{itemize}
\item  We propose GLIMG\footnote{A preliminary report of our work was accepted at IEEE BigData' 2019 \cite{lin2019fast}. We have extended it in the following aspects. $(1)$ By combining the global and local item graphs competently, we validate the effectiveness of our proposed GLIMG. (2) We discuss the benefit of taking into account the global taste and local preferences in this paper. (3) We conduct extensive experiments to analyze the sensitivity of hyper-parameters.}, a novel graph-based model which combines a global and multiple local item graphs to capture not only global preferences but also the more nuanced tastes within subgroups.
\item  By integrating item graphs into an adapted semi-supervised learning model, GLIMG encodes users' preferences and generate personalized recommendations.
\item We conduct extensive experiments and ablation studies to show the effectiveness of GLIMG method on two real-word datasets.
\end{itemize}

The rest of the paper is organized as follows. Related works are reviewed in section 2. We then detail our method GLIMG in section 3 and section 4. Intensive experiments are conducted in section 5. Finally, we conclude our work in section 6.

\section{Related Works}
There are two categories of works related to our work. In this section, we first study the graph-based recommendation methods and then review the different local models for recommendation.

\subsection{Graph-based recommendation}
Graph structures are widely used in the various recommendation tasks such as point-of-interest (POI) recommendation \cite{poi,zhang2020modeling}, web pages ranking \cite{pagerank}, online mobile recommendation\cite{he2020contextual,he2020learning}, and product recommendation \cite{itemrank,sun2019research,qiu2018bprh,sha2019attentive} for their natural modeling the relationship between entities.
A commonly studied case in recommender system is how to generate meaningful recommendation by exploiting limited user feedbacks. To alleviate such problem, different graph models are proposed to improve the performance of recommender systems. For example, Gu et al. propose GWNMF \cite{GWNMF}, a graph regularized nonnegative matrix factorization model, which constructs user and item graph, encoding the graph structure information or content information to regularize the latent factor model. Kang et al. \cite{kz2016} propose to preserve affinity and structure information about rating matrix by constructing both user and item graphs. He et al. \cite{birank, trirank} treat recommendation task as a vertex ranking problem. They propose BiRank \cite{birank} to model users and items as a user-item bipartite graph, which takes into account both the graph structure and the prior knowledge. TriRank \cite{trirank} allows additional information to be incorporated, which exploits review information and models the user-item-aspect as a tripartite graph. In our work, we extend the traditional item graph model by combining the global and local item correlations while taking into account the structural smoothness, which has not been studied before.

\subsection{Local models for recommendation}
Local models have been shown to achieve impressive performance in recommender systems \cite{focuslearning, LLORMA, clusteringitem}. Early works \cite{clusteringitem, scalable} cluster users or items and build multiple local models in order to improve the scalability of recommender systems.
Lee et al. \cite{LCR,LLORMA} assume the rating matrix is locally low-rank. LLORMA \cite{LLORMA} first choose anchor points and identify neighborhoods of each anchor point. A local low-rank latent factor model is estimated for every neighborhood and ratings are predicted by weighting all local models. LCR \cite{LCR} is then proposed to generalize LLORMA to a ranked loss minimization method. Their methods use local low-rank models to generate personalized recommendations, but they have not considered the global behavior of all users.

Christakopoulou et al. propose GLSLIM method \cite{glslim} which extends the SLIM model \cite{slim} by combining global and multiple local item-item models. They first use SLIM for estimating an item-item coefficient matrix, then separate the users into subsets and compute local item-item correlation matrix, allowing the user subsets to be updated. Since their method is based on SLIM, they only consider items that have been co-rated and thus fail to capture the transitive relationship between items that have not co-rated information \cite{fism}. A similar work GLSVD \cite{glsvd} is proposed as a latent approach to personalized combine global common aspects and local user subset specific aspects. Different from the above two works, our work is a graph-based approach, which propagates users' preferences to items based on their proximity. Besides, our method do not update the assignment of user subsets due to the unsatisfied efficiency \cite{glslim,glsvd} while our method outperforms competing top-N recommendation methods.

\section{Recommendation on Item Graph}
In this section, we first introduce the main mathematical notations used in this paper. Then we describe how to construct the item graph in detail. We also present two related hypotheses for item graph (i.e. rating smoothness and a soft constraint for the learned ratings).
\subsection{Notation}
Let $G = (P, E) $ be an undirected item graph, where $P$ represents the set of item nodes, and $E$ represents the set of edges. We use $u$ and $p$ to represent the users and items, respectively. Edges carry the relationship strength between items. The relationship between item $i$ and item $j$ is depicted by a correlation score $w_{ij}$. We denote the number of users by $m$ and the number of items by $n$. The rating matrix is denoted by $R$ of size $m \times n$, we use $R_u$ and $R_i$ to represent the observed rating of users and items, respectively. Predicted values are denoted by a \texttt{\char`\~}  over it, for example, the final predicted rating matrix is represented as $ \Tilde{R}$.
\subsection{Item Graph}
Current recommender systems face the data sparsity problem and the performance of most recommendation methods could degrade significantly when there is insufficient rating data. For example, a large amount of labeled rating data is needed for well trained deep learning models on recommendation tasks. The data sparsity problem motivates the semi-supervised learning (SSL) techniques to utilize the unlabeled data and alleviate the sparseness of recommendation datasets \cite{poi}.
Graph-based SSL, as one of SSL techniques, performs label propagation on the affinity graphs, and then the label information (rating) propagates from labeled data (e.g. items with rating) to unlabeled data (e.g. items without rating) based on their proximity, which is well-suited for the real-world recommender systems \cite{HF}. In graph-based recommendation methods, the entities such as users or items are represented as nodes, and the interactions or similarities between the nodes are encoded by edges.
Inspired by item-item models which work well on top-N recommendation task for their satisfactory performance and high scalability, in this work, we construct item graphs by merging users' preferences into the item nodes. The entries of each item node $p_i \in \mathbb{R}^m$ is the $i$th column of rating matrix R.
To construct an item graph, the edges which encode the proximity information between nodes are necessary to estimate. There are different similarity measurements such as cosine similarity and Pearson correlation coefficient \cite{itemcf}. For example, the cosine similarity between item i and item j, is computed by:
\begin{equation}
cos(p_i,p_j)=\frac{p_i\cdot p_j}{\|p_i\|\|p_j\|}
\end{equation}
where ` $\cdot$ ' represents the vector dot product operation and $\|\cdot\|$ denotes the Euclidean norm. However, the similarity score could be negative since the output of the cosine similarity is [-1,1]. To avoid such case, in this work, the correlation between a pair of items $p_i$ and $p_j$ are projected to the range [0,1] and computed by:
\begin{equation}
w_{ij}=exp\{-\sigma[1-cos(p_i, p_j)]\}
\end{equation}
where $\sigma$ is a hyper-parameter, and $cos(p_i, p_j)$ denotes the cosine similarity between item nodes.

\subsection{Rating Smoothness}
Users usually tend to give similar ratings on similar items. To capture the above assumption and keep the predicting scores be smooth enough with respect to the intrinsic structure of the item nodes \cite{itemgraph,lgc}, the graph regularization is formulated as follows:
\begin{equation}
\begin{split}
\min_{\Tilde{R}} &\frac{1}{2}\sum_{i,j=1}^{n} W_{ij}\| \frac{ \Tilde{R_i}}{\sqrt{D_{ii}}}-\frac{ \Tilde{R_j}}{\sqrt{D_{jj}}}\|^2 \\
\end{split}
\end{equation}
where $ \Tilde{R_i}$ and $ \Tilde{R_j}$ represent the final users' preference scores which need to be learned by the model. $\Tilde{R_i}$ is a vector whose dimension is $m$ , representing the predicted ratings of all users to item $i$. $W_{ij}=[w_{ij}]$ represents the all edge weight of the item graph, calculated by Eq. (2). In addition, $D_{ii}=\sum_{j} W_{ij}$ is a diagonal matrix which encodes the degree of nodes for normalization.

\subsection{Hard Constraint}
In order to keep the observed ratings unchanged while assign similar predicting scores to similar items, a hard constraint is applied to the Eq. (3) and formulated as follows:
\begin{equation}
\begin{split}
\min_{\Tilde{R}} \ &\frac{1}{2}\sum_{i,j=1}^{n} W_{ij}\| \frac{ \Tilde{R_i}}{\sqrt{D_{ii}}}-\frac{ \Tilde{R_j}}{\sqrt{D_{jj}}}\|^2 \\
s.t. \ &\Tilde{R_i}(rated) = R_i\\
\end{split}
\end{equation}
where $\Tilde{R_i}(rated)$ means the observed rated scores of the predicted item vector.

\subsection{Soft Constraint}
However, to solve a hard constraint is usually computational inefficient \cite{itemgraph}. Following \cite{itemgraph}, the hard constraint in Eq. (4) is replaced with a soft one through introducing a regularization term.
\begin{equation}
\begin{split}
\min_{\Tilde{R}} \ &\frac{1}{2}\sum_{i,j=1}^{n} W_{ij}\| \frac{ \Tilde{R_i}}{\sqrt{D_{ii}}}-\frac{ \Tilde{R_j}}{\sqrt{D_{jj}}}\|^2 +\mu \sum_{i=1}^n\| \Tilde{R_i}-R_i\|^2
\end{split}
\end{equation}
where $\mu$ denotes the weight of regularization term.

\section{The GLIMG approach}
In this section, we first introduce our proposed GLIMG algorithm, then we provide a close-form solution for GLIMG. At last, we analyze the time complexity of GLIMG.
\subsection{Global and Local Item Graphs (GLIMG)}
In this work, we argue that a single global item graph shared by all users can not accurately capture the diverse tastes between different user subgroups. We tackle the above problem by constructing multiple local items graphs along with a global item graph, in this way, our model GLIMG simultaneously captures the global and local user tastes.

To be specific, we first assign all users into $k$ different subgroups using clustering algorithms. In this work, we use a popular clustering approach k-means++ algorithm \cite{kmeanspp} in the experiments and it is conceivable that more adaptive clustering algorithms can lead the model to achieve better performance. As a variant of k-means, the process of the k-means++ method is as follows:
\begin{enumerate}
\item A data point is randomly chosen as the first center.
\item $k-1$ points are iteratively selected as the centers based on the distance from data points to the nearest center that has already been chosen.
\item Assign the set of data points to the $k$th cluster if they are closer to the $k$th center than other centers.
\item Recalculate the center point of each cluster.
\item Repeat step 3 and step 4 until convergence.
\end{enumerate}

Similar to GLSLIM \cite{glslim}, we then split the user-item rating matrix $R$ into $k$ training matrices as user subgroups which are denoted as $R^{L_u}$, $\forall L_u \in \{1,...,k\}$.
After assigning users into $k$ subsets, there are $k$ training matrices which are denoted as $R^{L_u}$, $\forall L_u \in \{1,...,k\}$. If the user u belongs to the $L_u$th subset, then the $u$th row of $R^{L_u}$ will be the $u$th row of $R$, otherwise, the $u$th row of $R^{L_u}$ will be empty. In other words, each user's rating information will be only appear in one training matrix.
We then estimate a local item correlation matrix $W^{L_u}$ $\forall L_u \in \{1,...,k\}$ for every user subgroup in order to combine the global and local item correlations.

To better model the item graphs and achieve personalized recommendation, we follow the traditional assumptions and design our model based on the following hypotheses.

\textbf{Hypothesis 1. Item Smoothness} The preference scores of the two items should be similar if the items are among similar graph structures.

\textbf{Hypothesis 2. Item Fitting} The item preference scores should not change too much from the initial assignment. For example, if an item receives higher rating scores than others, it is likely to receive more high ratings in the future.

\textbf{Hypothesis 3. Item Confidence } In collaborative filtering, two items are considered to have similarity if they are purchased by the same users. In other words, the similarity of the two items can be estimated by the number of users they interacted with. However, popular items can cause inaccuracy similarity calculation between items, which can result in the final recommendation results not consistent with the users' preferences. For example, Amazon researchers found that anyone who has bought books seems to have bought "Harry Potter", which makes many books are related to "Harry Potter". An underlying reason was the popularity of "Harry Potter", such behavior of purchasing this book is not useful for predicting users' preferences \cite{smith2017two,gao2018similarity}.
In addition, popular items are more likely to be viewed or known by the user, thus the absence of an interaction with these items is more likely to be treated as negative feedback in recent works \cite{gao2018similarity,steck2011item,wu2019noise}. Such phenomenon is called "item popularity bias" \cite{chen2020bias}. To debias the item popularity and achieve personalized recommendation, we design a regularization term called "item confidence".
If an item is highly correlated to many other items, the confidence of this item should be small and in order to generate personalized recommendation for users, this item's preference scores should be suppressed.

To capture the above three hypotheses, we devise the following graph regularization function for every user:

\begin{equation}
\footnotesize
\begin{split}
\min_{\Tilde{R}} \ \frac{1}{2}\sum_{i,j=1}^{n} ( g W_{ij}+ (1-g)  W_{ij}^{L_u})\| \frac{ \Tilde{R_i}}{\sqrt{D_{ii}}}-\frac{ \Tilde{R_j}}{\sqrt{D_{jj}}}\|^2 + \mu \sum_{i=1}^n\| \Tilde{R_i}-R_i\|^2+ \gamma \sum_{i=1}^{n}\|\sqrt{D_{ii}} \Tilde{R_i}\|^2
\end{split}
\end{equation}
where hyper-parameter $g$ balances the effect of global and local item correlation, the regularization parameters $\mu$ and $\gamma$ control the weight among these three terms. We also constrain $W_{ii}=0$ and $W_{ii}^{L_u}=0$ to prohibit the item from recommending itself.

The first term in the Eq. (6) is devised to capture hypothesis 1, and the weighted normalization scheme lets the item nodes consider its local graph structure during the propagation. The second term is devised to capture hypothesis 2, ensuring the global consistency between item nodes.
In addition, we propose the third term of Eq. (6) to capture hypothesis 3, which restrain the "common choices" and lead the model to achieve personalized recommendation. In order to verify the effectiveness of this regularization term, we also conduct sensitivity experiments of $\gamma$ in section 5.4, showing the importance of the third regularization term in achieving the personalized recommendation.

\subsection{Model Optimization}
In order to solve Eq. (6), we first introduce a normalized correlation matrix $S$ defined as:
\begin{equation}
S_{ij}^{L_u}=\frac{gW_{ij}+(1-g)W_{ij}^{L_u}}{\sqrt{D_{ii}}\sqrt{D_{jj}}}
\end{equation}
Then the Eq. (6) can be written as:

\begin{equation*}
\min_{\Tilde{R}} \sum_{i,j=1}^{n}\Tilde{R_i}^T \Tilde{R_j} - \sum_{i,j=1}^{n}\Tilde{R_i}^T \Tilde{R_j} S_{ij}^{L_u}+ \mu \sum_{i=1}^n\| \Tilde{R_i}-R_i\|^2+\gamma \sum_{i=1}^{n}D_{ii} \|\Tilde{R_i}\|^2 \\
\end{equation*}
We set the derivative of Eq. (6) with respective to $\Tilde{R}$ to 0. Then we denote the optimal solution of the above objective function as $\Tilde{R}^{L_u}$ and obtain:
\begin{equation*}
\begin{split}
 \Tilde{R}^{L_u}- \Tilde{R}^{L_u}S^{L_u}+ \mu( \Tilde{R}^{L_u}-R^{L_u})+\gamma  \Tilde{R}^{L_u}D=0 \\
 \Tilde{R}^{L_u}-\frac{1}{1+\mu} \Tilde{R}^{L_u}S^{L_u}-\frac{\mu}{1+\mu}R^{L_u}+\frac{1}{1+\mu}\gamma  \Tilde{R}^{L_u}D=0 \\
 \Tilde{R}^{L_u}-\alpha  \Tilde{R}^{L_u}S^{L_u}-\beta R^{L_u}+\alpha \gamma  \Tilde{R}^{L_u}D=0 \\
\end{split}
\end{equation*}
where $\alpha=\frac{1}{1+\mu}$, $\beta=\frac{\mu}{1+\mu}$. Finally, the closed-form solution can be easily derived:
\begin{equation}
 \Tilde{R}^{L_u} = \beta R^{L_u}[I+\alpha(\gamma D-S^{L_u})]^{-1}
\end{equation}
where $D_{ii}=\sum_{j} S_{ij}^{L_u}$, and $I$ is an identity matrix.
Since we focus on the top-N recommendation task, $\beta$ could be ignored as it is shared by all the items. In the meantime, we set the $\gamma=1$ and $L=D-S$ is the graph Laplacian \cite{gl} of the item graph. Therefore, the Eq. (8) can be rewritten as:
\begin{equation}
   \Tilde{R}^{L_u} =  R^{L_u}(I+\alpha L^{L_u})^{-1}
\end{equation}
In this way, the missing ratings for all user subsets can be estimated by Eq. (9) with corresponding training matrix and item correlations.

\subsection{An Iterative Solution}

One may wonder whether our proposed method can be learned in an iterative way. To answer the above question, we formulate the iterative algorithm as follows:

\begin{itemize}
\item Compute the global item-item similarity with $W_{ij}=exp\{-\sigma[1-cos(p_i, p_j)]\}$ \;
\item Cluster users with k-means++ algorithm and compute local item-item similarity $\forall L_u \in \{1,...,k\}$: $W_{ij}^{L_u}=exp\{-\sigma[1-cos(p_i, p_j)]\}$ \;
\item Compute the graph Laplacian: $L^{L_u}=D-S^{L_u}$, where $S_{ij}^{L_u}=\frac{gW_{ij}+(1-g)W_{ij}^{L_u}}{\sqrt{D_{ii}}\sqrt{D_{jj}}}$, and $D_{ii}=\sum_{j} S_{ij}^{L_u}$ \;
\item Iterate $\Tilde{R}^{L_u}(t+1)= R^{L_u}-\alpha \Tilde{R}^{L_u}(t) L^{L_u}$ until convergence.
\end{itemize}

To show the convergence of our iterative algorithm, we have $\Tilde{R}(t+1) = \Tilde{R}(t)$. Hence
\begin{equation}
\begin{split}
\Tilde{R}^{L_u}(t+1)= R^{L_u}-\alpha \Tilde{R}^{L_u}(t) L^{L_u} \\
\Tilde{R}^{{L_u}^{*}} + \alpha  \Tilde{R}^{{L_u}^{*}} L^{L_u} =  R^{L_u}\\
\Tilde{R}^{{L_u}^{*}} (I+\alpha L^{L_u})= R^{L_u} \\
\Tilde{R}^{{L_u}^{*}} = R^{L_u} (I+\alpha L^{L_u})^{-1}
\end{split}
\end{equation}
We show that the optimal solution obtained by the above iterative optimization algorithm is exactly the same as our proposed closed-form solution in Eq. (9). This suggested that our closed-form solution is the optimal solution, and the iterative algorithm would not be necessary.

\subsection{Online Recommendation and Time Complexity Analysis}
In this subsection, we describe how GLIMG works and analyze its time complexity. The overview of GLIMG is given in Algorithm 1. As shown in Algorithm 1, the first step is to compute the global item correlation matrix, whose time complexity is $O(mn^2)$. We then cluster the users into different user subgroups by using k-means++ method, and the time complexity of this step is $O(tkmn)$ where $t$ denotes the number of iterations, and $k$ is the number of clusters. For each subgroup of users, we then compute the local item correlation matrices, and the time complexity is $O(kmn^2)$. After that, we obtain the final item  correlation matrix for each user by combining the global item correlation matrix and the user's corresponding local item correlation matrix, and the time complexity of this step is $O(kn^2)$. At last, we compute the inverse of matrix $(I+\alpha L)$, whose time complexity is $O(kn^3)$. 
Furthermore, It is also worth noting that the above processes can be completed \textbf{offline}.

For an active user, we first retrieve her historical ratings and record the ratings she has made. Then the online personalized recommendation can be generated by ranking items according to their user preference scores which are predicted by Eq. (9). In this way, the top-ranked items are recommended to this active user. Since we only need to compute a vector-matrix multiplication online, the online computational complexity is only \textbf{$O(n)$}. 
Note that GLIMG is able to generate meaningful recommendations to users who have provided relative small number of ratings (i.e. the cold start users). Even in extreme cases, our proposed algorithm can make recommendations after a completely  new user has made one rating.

\begin{algorithm}[t]
	\SetAlgoLined
	\SetKwInput{offline}{Offline Training}
	\SetKwInput{online}{Online Recommendation}
	\KwIn{Rating matrix $R$, number of clusters $k$, \newline$\mu$ which controls the fitting constraint,\newline $\sigma$ which controls the similarity measurement, \newline $g$ which controls the balance between global and local models }
	\KwOut{Target user's preference scores for items $ \Tilde{R_u}$}
	\offline {for all users}
	Compute the global item correlation matrix with Eq. (2): $W_{ij}=exp\{-\sigma[1-cos(p_i, p_j)]\}$\;
	Cluster users with k-means++ algorithm\;
	\For {all clusters $L_u$}{
		Compute the local item correlation matrices with Eq. (2) $\forall L_u \in \{1,...,k\}$: $W_{ij}^{L_u}=exp\{-\sigma[1-cos(p_i, p_j)]\}$ \;
		Combine the global item correlation matrix and corresponding local item correlation matrix: $S_{ij}^{L_u}=\frac{gW_{ij}+(1-g)W_{ij}^{L_u}}{\sqrt{D_{ii}}\sqrt{D_{jj}}}$ \;
		Compute the graph Laplacian
		$L^{L_u}=D-S^{L_u}$ and the inverse matrix $(I+\alpha L^{L_u})^{-1}$, where $D_{ii}=\sum_{j} S_{ij}^{L_u}$
		
	}

	\online{for the target user u}
	Assign the user to the corresponding local group\;
	Estimate the missing ratings by Eq. (9) with her historical ratings and corresponding item correlations: $ \Tilde{R_u} =  R_{u}(I+\alpha L^{L_u})^{-1} $
	\caption{GLIMG}
\end{algorithm}

\section{Experiments and Analysis}
In this section, we conduct extensive experiments and ablation studies to show the effectiveness of our method.\\
\subsection{Experimental Setup}
\textbf{Datasets.} We evaluate our proposed method on two public datasets whose statistics are shown in Table 1. The first dataset, MovieLens-1M \footnote{https://grouplens.org/datasets/movielens} is a famous movie dataset, which contains 1M rating information with 6,040 users and 3,706 movies. The second dataset is Yelp. It concludes over 5M ratings with 1,326,101 users and 174,567 items. This dataset is about users' preference on restaurants, released by Yelp as Yelp Challenge Dataset on January 2018 \footnote{https://www.yelp.com/dataset\_challenge}, which is very sparse. Similar to \cite{glslim,trirank}, we create the subset by keeping users and items that have at least 30 ratings. All the ratings range from 1 to 5. Note that GLSVD can only deal with implicit feedback. Therefore, we follow \cite{glsvd} to transform above 2 datasets into implicit data for GLSVD approach, where each entry is marked as 0 or 1 indicating whether the user has rated the item.
\begin{table}[h]
	\centering
	\
	\begin{tabular}{ccccc}
		\hline
		Datasets     & \#User & \#Item & \#Rating  & Density \\ \hline
		MovieLens-1m & 6,040   & 3,706   & 1,000,209 & 4.24\%  \\
		Yelp 2018         & 5,684   & 4,961   & 356,889  & 1.26\%  \\ \hline
	\end{tabular}
	\caption{ Statistics of datasets}
\end{table}
\\
\textbf{Evaluation methodology.} We split each dataset into 3 parts, the first 80\% are used for training, 10\% are used for validation and the remaining 10
\% are used for testing the performance.
\\We use 4 evaluation measures which have been widely used in top-N evaluation. \textsl {Hit Ratio} (HR) shows that the percentage of users that have at least one correct recommendation \cite{glslim,zhang17}.
HR is defined as
\begin{equation}
HR@N=\frac{\text{Number of hits}}{\text{Number of users}}
\end{equation}
\textsl {Normalized Discounted Cumulative Gain} (NDCG) considers the position of correct recommendations \cite{zhang17,sun2020we}. The score is averaged across all the testing users. To be specific, $Z_N$ is a normalizer to guarantee the perfect recommendation score is 1. The relevance score of item at position is denoted as $r_i$, $r_i$=1 if the recommended item is in the test set and 0 otherwise \cite{trirank}.
\begin{equation}
NDCG@N=Z_N\sum_{i=1}^{N} \frac{2^ {r_i} -1}{log_2(i+1)}
\end {equation}
Precision measures the percentages of correct recommendations and also averaged across all testing users.
\begin{equation}
\small
Precision@N={\frac{\text{Number of items the user likes among the top-N recommendations} }{|N|}}
\end{equation}
Recall evaluates the percentage of purchased items that are in the recommendation list, averaged across all testing users.
\begin{equation}
\small
Recall@N={\frac{\text{Number of items the user likes among the top-N recommendations} }{\text{Total number of items the user likes}}}
\end{equation}\\

\textbf{Comparison algorithms.}
\begin{itemize}
  \item \textbf{Item Popularity (ItemPop)} Items are ranked by their popularity which is the number of ratings. This method is not personalized and thus can be seen as the baseline \cite{performance2010}.
  \item \textbf{ItemRank} \cite{itemrank} This is a graph-based method for top-N recommendation, which predicts users' preferences according to item-item correlation graphs. We follow the implementation in \cite{trirank}.
  \item \textbf{SLIM} \cite{slim} This is a state-of -the-art for top-N recommendation, which learns item-item aggregation coefficient. We use LibRec library \cite{librec} to conduct the experiments.
  \item \textbf{Recommendation on Dual Graphs (RODG)} \cite{kz2016} This is also a graph-based method for top-N recommendation. This method adopts graph regularization to incorporate user graph and item graph. The source code is available \footnote{https://github.com/sckangz}.
  \item \textbf{BiRank} \cite{birank} BiRank ranks vertices of bipartite (user-item) graphs and generates meaningful recommendations. We follow the implementation in \cite{birank}.
  \item \textbf{GLSVD}\cite{glsvd} As a state-of-the-art global-local model, GLSVD captures the global and local user tastes by a latent space approach. We choose rGLSVD approach as a benchmark and follow the implementation in \cite{glsvd}. The source code is available\footnote{https://github.com/echristakopoulou/glsvd}.
\end{itemize}

\textbf{Model Selection.} In order to find the optimal parameters which lead to best performance of each methods, we hence conduct extensive search over the parameter space of each method. For ItemRank, we use the common choice for the value of decay factor $\alpha$ which is 0.85 \cite{itemrank}. For SLIM, the value of the \textsl{$l_1$} and \textsl{$l_2$} regularization parameters we tried are $\lbrace$0.001, 0.01, 0.1, 1, 3, 5, 7, 10, 15, 20, 30, 50, 80$\rbrace$. For RODG, we select the value of $\alpha$ and $\beta$ from the following sets \cite{kz2016}: $\{1e^{-6},1e^{-4},1e^{-2}\}$ and $\{1e^{-6},1e^{-5},1e^{-4},1e^{-3},1e^{-2}\}$, respectively. For BiRank, we apply grid search in the following set: $\lbrace$0.001, 0.1, 1, 3, 5, 7, 10, 15, 20, 30$\rbrace$ to find the optimal regularization parameters $\gamma$ and $\eta$. As for rGLSVD, the number of clusters $Cls$ is chosen from the following set: $\{3,5,8,10,15,20,25,30,35,40,45,50,$ $70,100\}$, and the rank of global model $f^g$ is selected from the following set $\{1, 3, 5, 10, 20, 30, 40, 50, 60, 70, 80, 90, 100\}$. In addition, we use CLUTO \cite{cluto} to cluster users for rGLSVD approach.

\subsection{Performance Comparison with Baselines}

\begin{table*}[t]
	\footnotesize
	\begin{adjustwidth}{-0.95in}{0in}
	\centering
	\caption{Comparison with competing methods at rank 10 and 50 (i.e.N).}
	\begin{tabular}{ccccc|cccc}  \hline
	Dataset    & \multicolumn{4}{c}{ML-1M}                                         & \multicolumn{4}{|c}{Yelp}                                       \\ \hline
	Metric(\%) & NDCG@10        & HR@10          & Precision@10   & Recall@10      & NDCG@10       & HR@10          & Precision@10  & Recall@10     \\ \hline
	ItemPop    & 11.03          & 50.75          & 8.97           & 7.04           & 1.73          & 10.03          & 1.04          & 2.00          \\ \hline
	ItemRank   & 13.79          & 57.99          & 10.79          & 8.97           & 5.09          & 27.66          & 3.08          & 5.87          \\ \hline
	SLIM       & 21.31          & 74.43          & 16.31          & 15.57          & 6.91          & 35.01          & 4.22          & 7.84          \\ \hline
	RODG       & 19.49          & 70.17          & 14.75          & 13.82          & 7.44          & 36.88          & 4.44          & 8.45          \\ \hline
	BiRank     & 17.52          & 67.23          & 13.03          & 12.42          & 6.93          & 34.92          & 4.10          & 7.90          \\ \hline
	GLSVD      & 19.62          & 73.43          & 14.82          & 15.49          & 5.64          & 30.20          & 3.42          & 6.74          \\ \hline
	GLIMG      & \textbf{23.08} & \textbf{78.42} & \textbf{17.08} & \textbf{17.73} & \textbf{7.81} & \textbf{37.37} & \textbf{4.60} & \textbf{8.76} \\ \hline
\end{tabular}

	\centering
	\begin{tabular}{ccccc|cccc}  \\ \hline
	Dataset    & \multicolumn{4}{c}{ML-1M}                                        & \multicolumn{4}{|c}{Yelp}                                  \\       \hline
	Metric(\%) & NDCG@50        & HR@50          & Precision@50  & Recall@50      & NDCG@50        & HR@50          & Precision@50  & Recall@50      \\ \hline
	ItemPop    & 13.84          & 76.79          & 5.7           & 19.64          & 3.06           & 26.45          & 0.70          & 5.81           \\ \hline
	ItemRank   & 16.85          & 81.34          & 6.66          & 23.21          & 8.92           & 57.43          & 2.03          & 16.92          \\ \hline
	SLIM       & 26.44          & 90.99          & 9.66          & 36.60          & 11.72          & 65.78          & 2.64          & 21.67          \\ \hline
	RODG       & 22.98          & 88.15          & 8.16          & 30.79          & 12.40          & 67.40          & 2.72          & 22.80          \\ \hline
	BiRank     & 21.96          & 88.45          & 7.92          & 30.94          & 11.88          & 66.51          & 2.64          & 22.10          \\ \hline
	GLSVD      & 25.51          & 91.31          & 9.11          & 36.63          & 9.88           & 62.11          & 2.19          & 18.81          \\ \hline
	GLIMG      & \textbf{28.66} & \textbf{93.77} & \textbf{9.90} & \textbf{40.17} & \textbf{13.20} & \textbf{70.27} & \textbf{2.89} & \textbf{24.27}  \\ \hline
\end{tabular}

	\end{adjustwidth}
\end{table*}
Table 2 shows the evaluated scores of competing methods and GLIMG at rank 10 and 50, while Figure 1 and Figure 2 plot the performance when the size of recommendation lists varies from 10 to 50. We first focus on the evaluations on the MovieLens dataset. As shown in Figure 1, all the metrics show the same trend: our method GLIMG achieves the best performance compared to other methods, followed by SLIM, GLSVD, RODG and BiRank. In addition, RODG outperforms BiRank when N is small, as evaluated by all metrics. When N is set to 40-50, the evaluated differences between RODG and BiRank are getting small except NDCG, which shows RODG is able to order the items more accurately than BiRank.
ItemRank is obviously worse than the above approaches.
ItemPop performs worst among the competing methods, which shows that only recommending popular items is not personalized enough to generate meaningful recommendation lists.

\begin{figure*}[!t]\vskip -0.05 in
	\begin{center}
		{\subfigure[MovieLens-NDCG]
			{\includegraphics[width=0.45\textwidth]{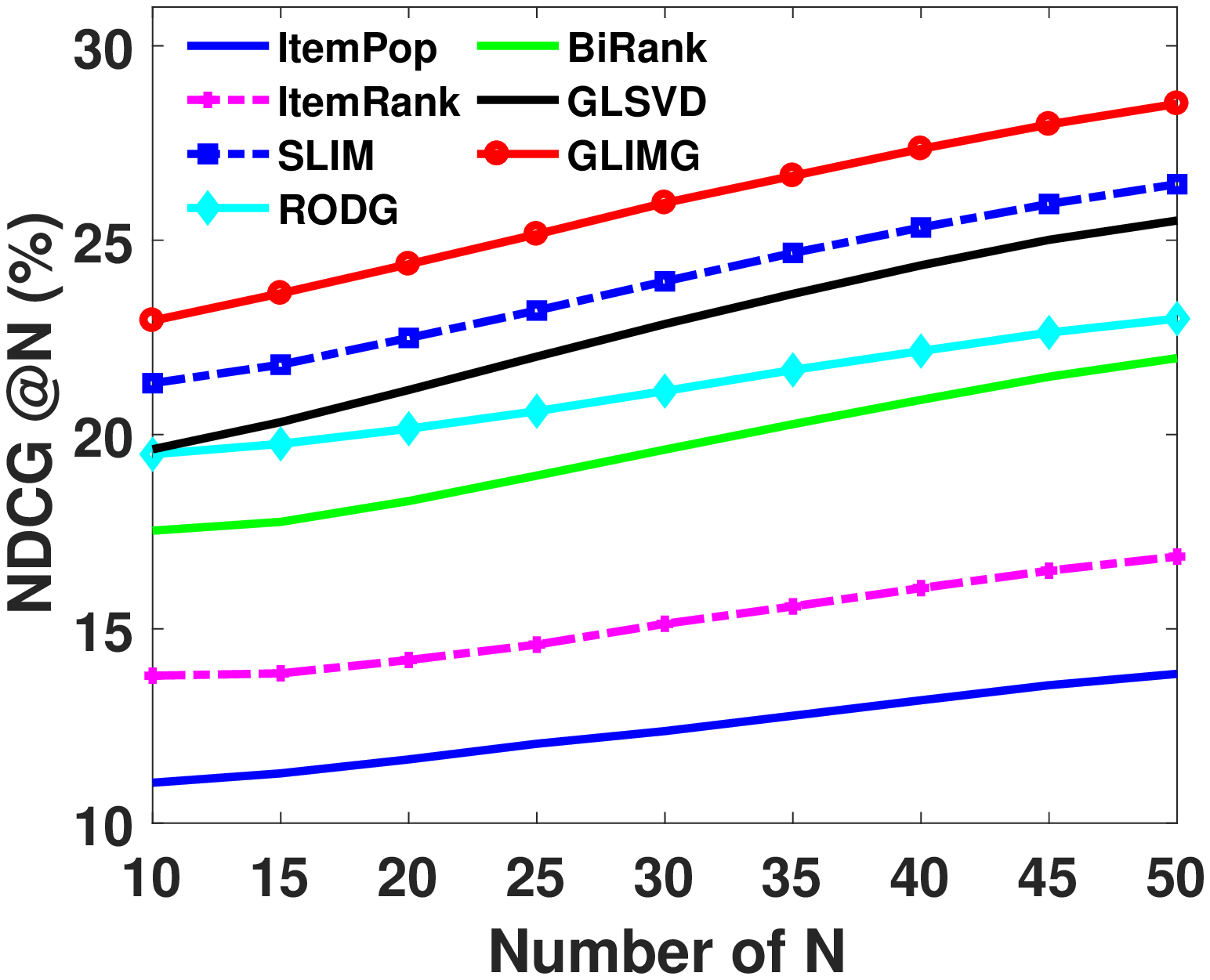}\label{fig2:a}}}
		{\subfigure[MovieLens-HR]
			{\includegraphics[width=0.45\textwidth]{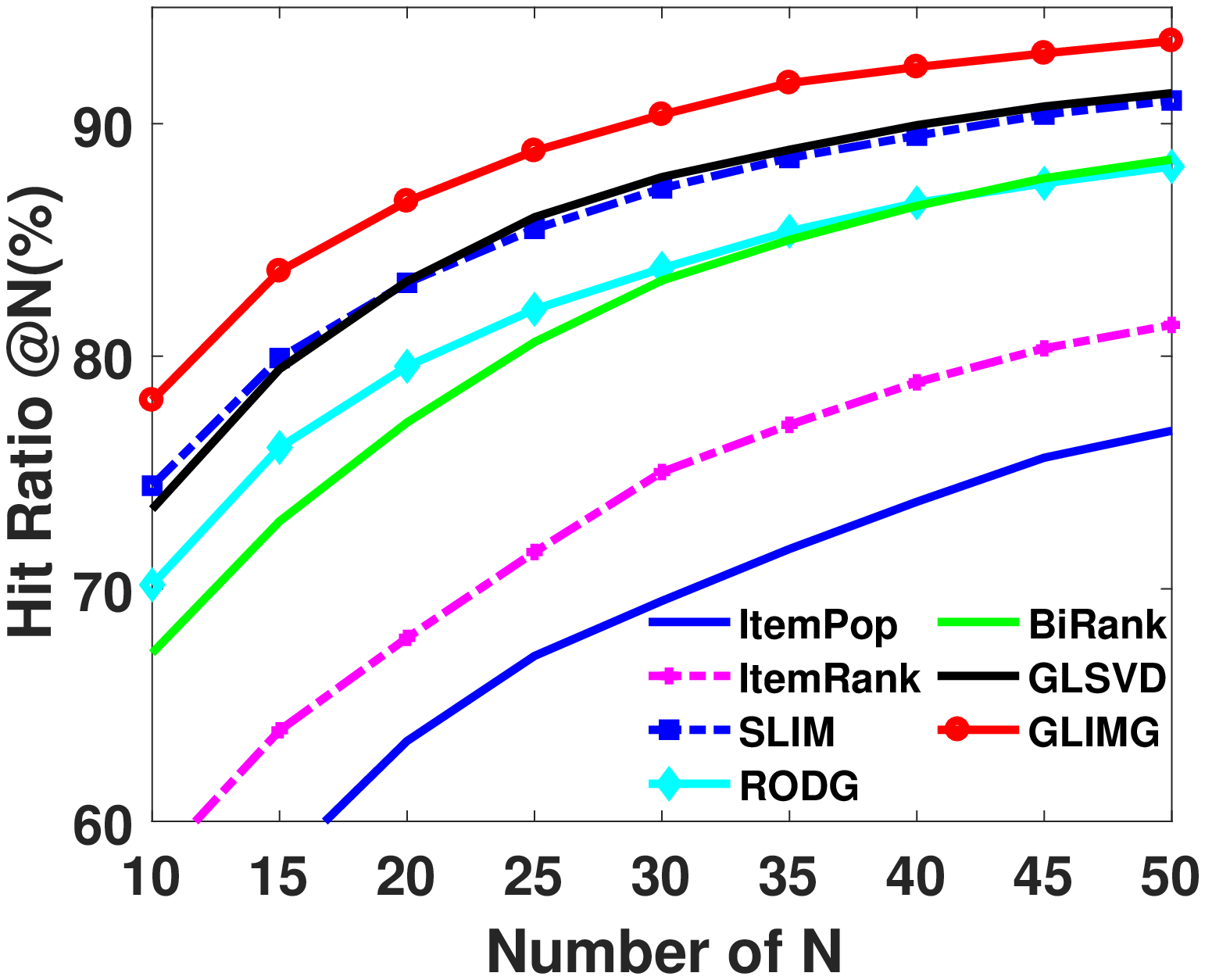}\label{fig2:b}}}
		{\subfigure[MovieLens-Precision]
			{\includegraphics[width=0.45\textwidth]{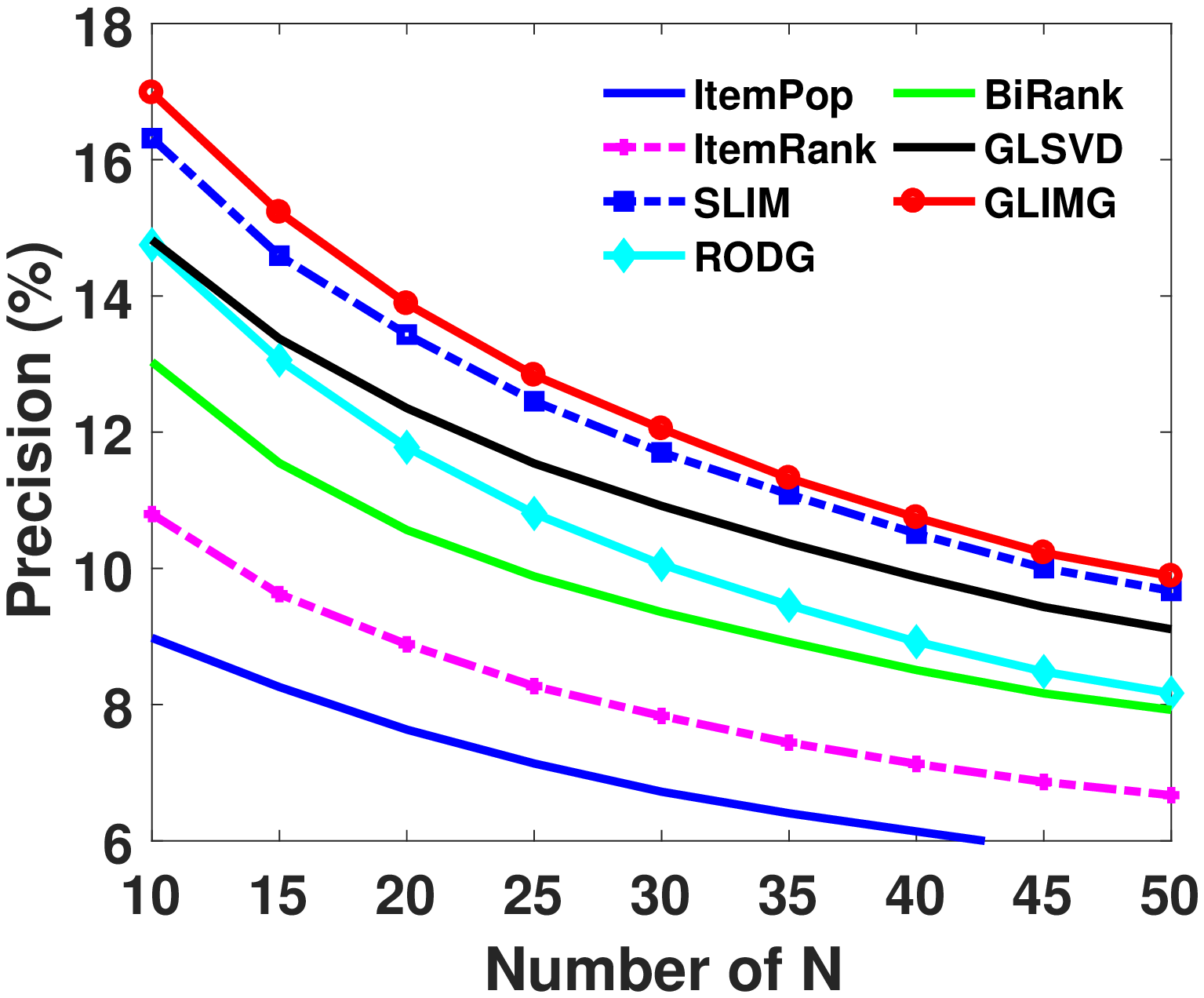}\label{fig2:c}}}
		{\subfigure[MovieLens-Recall]
			{\includegraphics[width=0.45\textwidth]{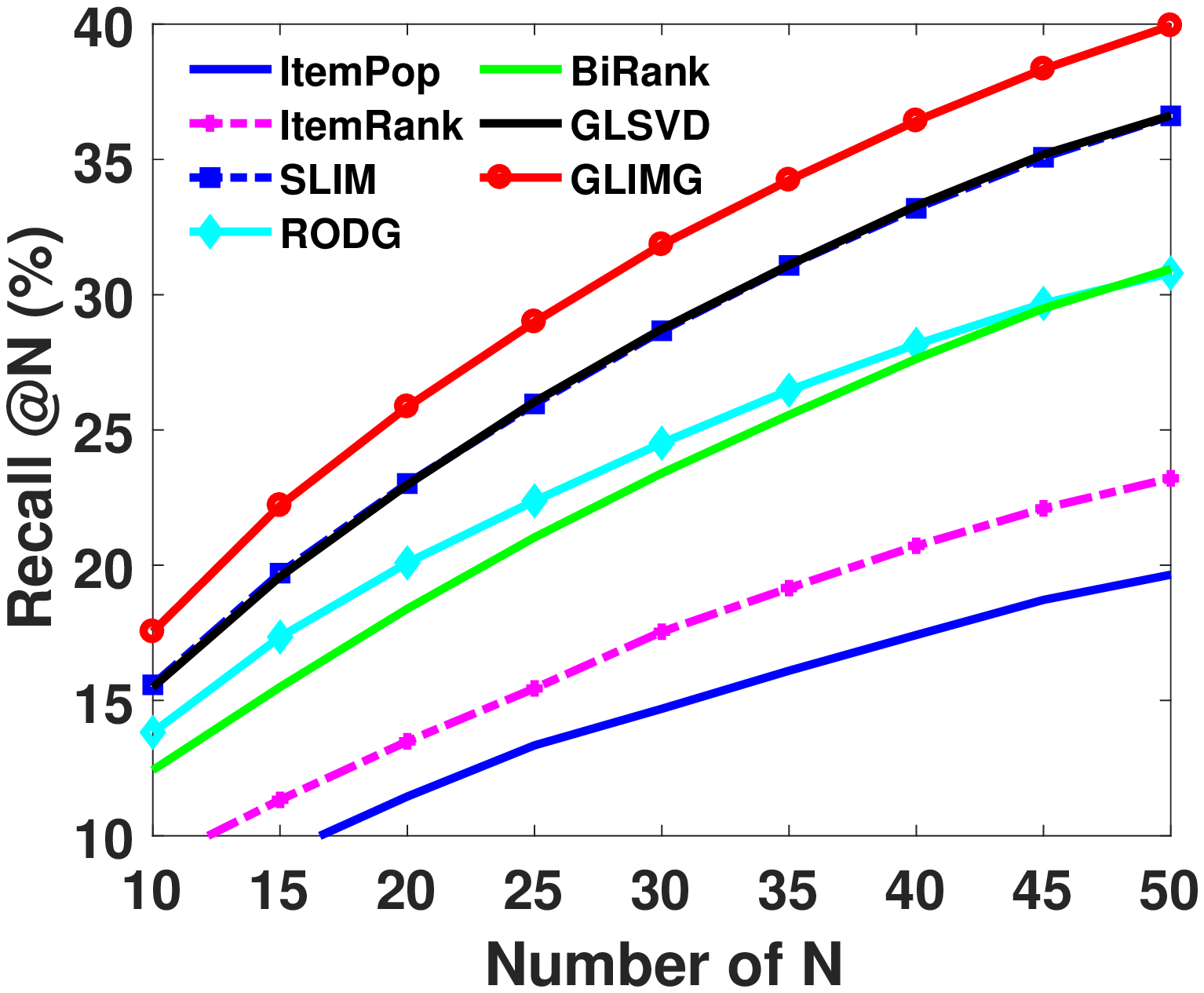}\label{fig2:d}}}
		\vskip -0.11 in
		\vspace*{8pt}
		\caption{Performance evaluated by different metrics from position 10 to 50 on MovieLens dataset.}
		\label{fig1}
	\end{center}\vskip -0.15 in
\end{figure*}

As for Yelp dataset, GLIMG again outperforms other competing approaches. Followed by RODG, which also significantly outperforms the rest methods.
BiRank and SLIM have a very similar performance evaluated by all metrics except by Recall. Specifically, SLIM slightly exceeds BiRank indicating that although SLIM can generate meaningful recommendations, it is unable to order the items very correctly. The cause of this phenomenon may be due to the missing item relationships that SLIM fails to capture.
Meanwhile, GLSVD and ItemRank performs poorly among the personalized recommendation methods.
ItemPop still achieves the worst performance and thus is omitted in Figure 3 in order to better display the performance of other approaches.

\begin{figure*}[!t]\vskip -0.05 in
	\begin{center}
		{\subfigure[Yelp-NDCG]
			{\includegraphics[width=0.45\textwidth]{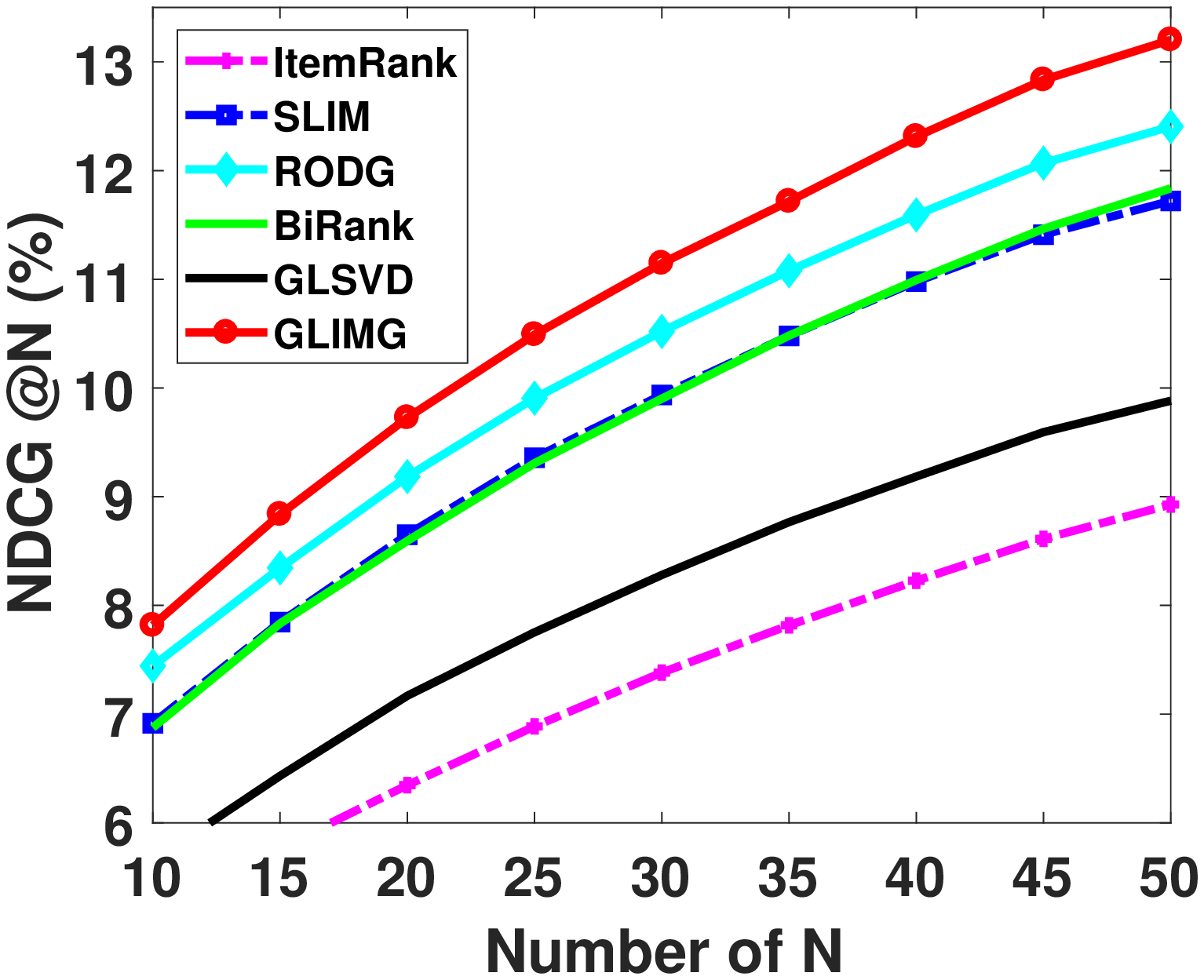}\label{fig3:a}}}
		{\subfigure[Yelp-HR]
			{\includegraphics[width=0.45\textwidth]{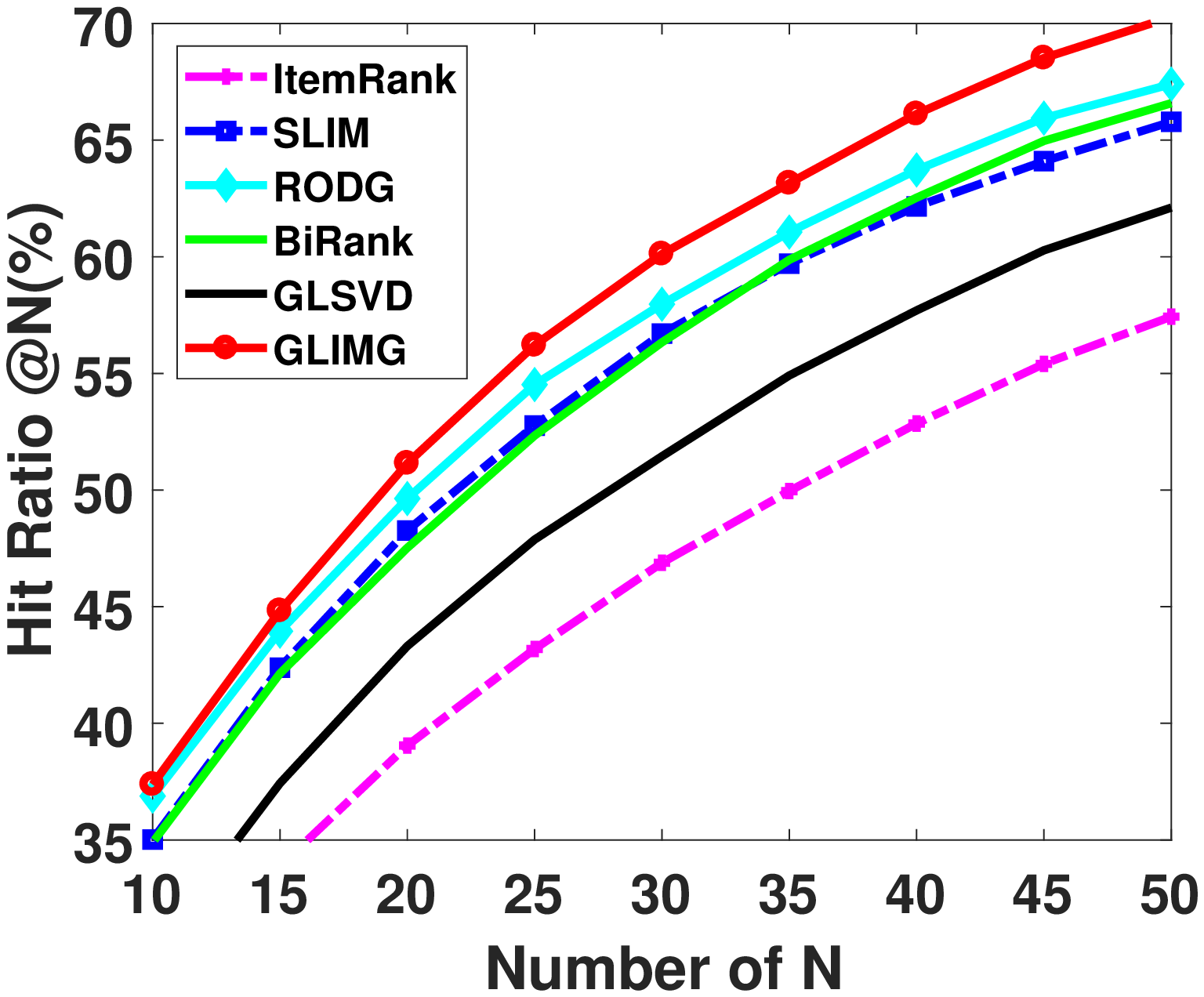}\label{fig3:b}}}
		{\subfigure[Yelp-Precision]
			{\includegraphics[width=0.45\textwidth]{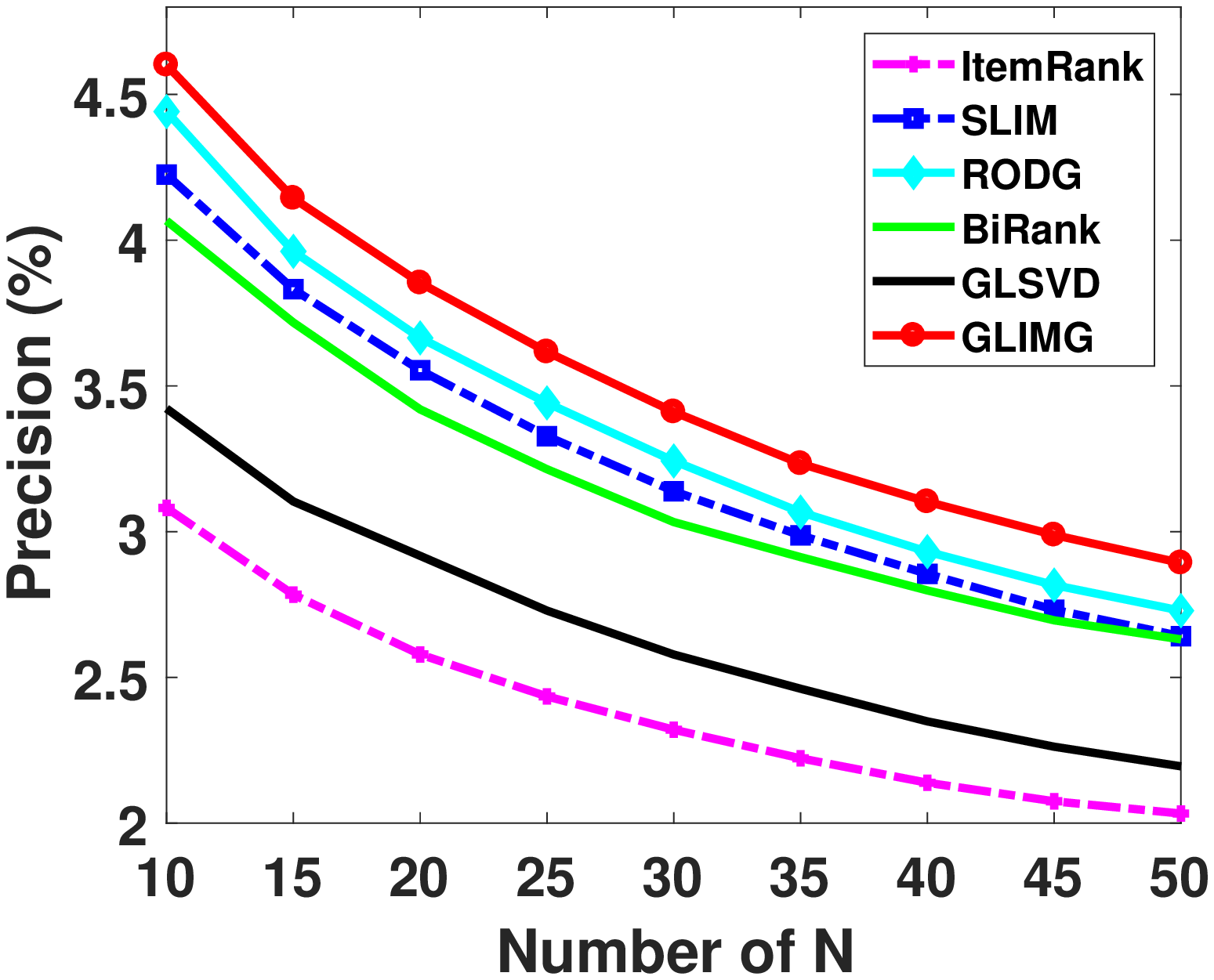}\label{fig3:c}}}
		{\subfigure[Yelp-Recall]
			{\includegraphics[width=0.45\textwidth]{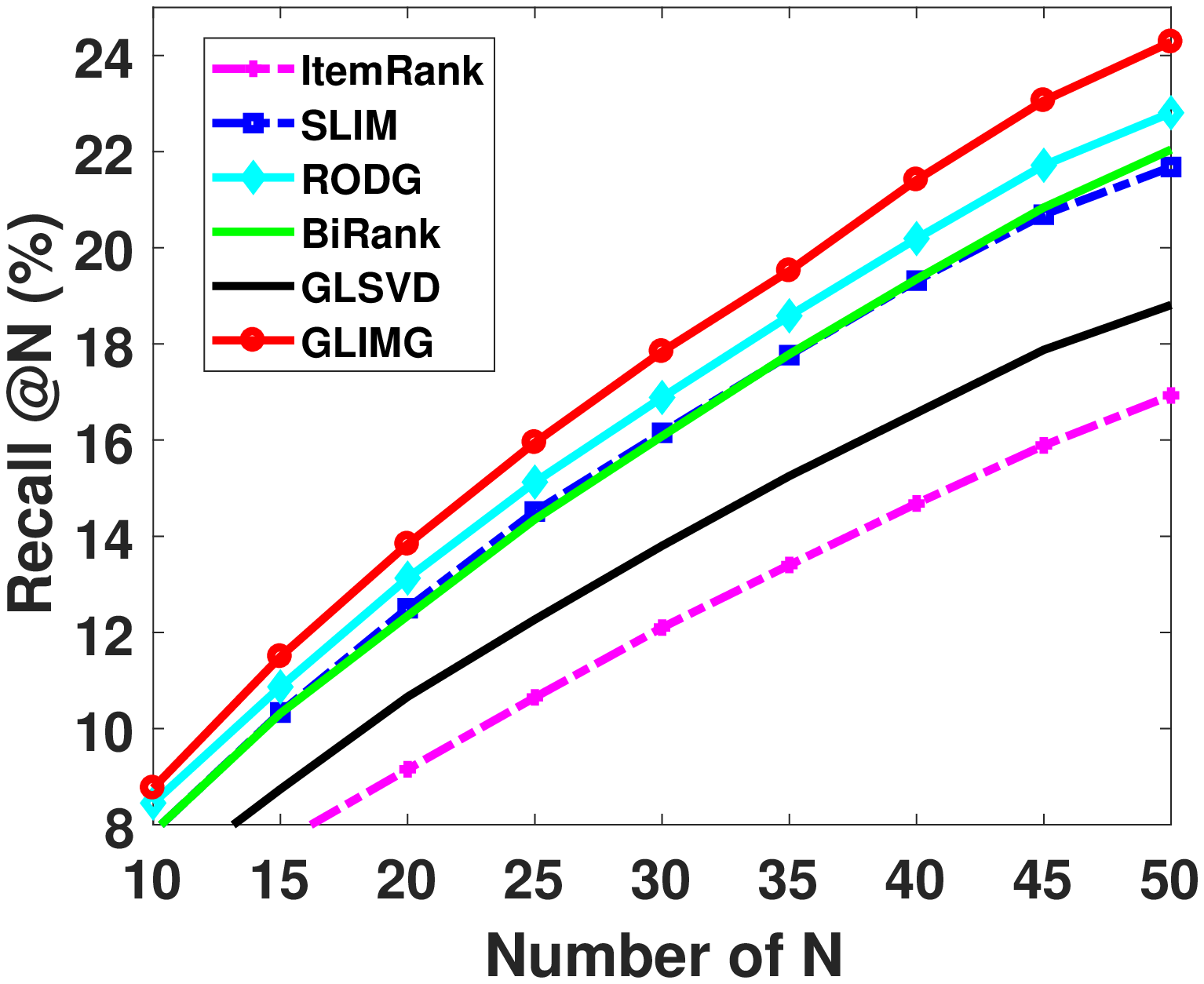}\label{fig3:d}}}
		\vskip -0.11 in
		\vspace*{8pt}
		\caption{Performance evaluated by different metrics from position 10 to 50 on Yelp dataset.}
		\label{fig2}
	\end{center}\vskip -0.15 in
\end{figure*}

Besides, there are some interesting findings across two datasets. Firstly, we notice that the ItemPop only performs well on the MovieLens dataset, the potential reason probably due to the two different domains of datasets: movies and restaurants, showing that users tend to watch popular movies online but unable to visit the popular restaurants due to the geographical reasons sometimes.
Another interesting finding is that the RODG and BiRank do not always underperform SLIM. For Yelp dataset, BiRank achieves competing performance compared with SLIM while RODG outperform two former methods. We believe that the underlying factor is the data sparsity problem. As shown in Table 1, since the density of Yelp dataset is only 1.26\%, the model-based methods face the difficulties to learn meaningful features from the rating matrix, while the graph-based methods enable to alleviate such problem by exploiting the potential relationships between entities.
In addition, although RODG and BiRank take into account the user and item information simultaneously, our model GLIMG which only exploits the item information, consistently outperforms than other graph-based methods. Overall, the superb performance of GLIMG shows the effectiveness of the proposed hypotheses and local graph models.

\subsection{Effectiveness of Global and Local Models}
Since our model contains the global and local parts, we also conduct extensive experiments on the following methods in order to investigate their impact on the overall recommendation:
\begin{itemize}
  \item \textbf{GIMG (\underline{G}lobal \underline{I}te\underline{M} \underline{G}raph)} No local models are estimated in this method, GIMG represents the global element of the GLIMG (i.e. $g=0$).
  \item \textbf{LIMG (\underline{L}ocal \underline{I}te\underline{M} \underline{G}raphs)} No global model is estimated in this method, LIMG is denoted as the local element of the GLIMG (i.e. $g=1$).
\end{itemize}

We organize the performance of our proposed models (i.e. GLIMG, GIMG, and LIMG) in Table 3 to verify the effectiveness of global and local models. It is obvious that GLIMG outperforms other two methods on both datasets, as evaluated by all metrics. The comparison strongly illustrates that such combination of global model and multiple local models can improve the recommendation quality without any additional information.

Figure 3 plots the NDCG scores with respect to the balance parameter between the global model and the local models, which validates the effectiveness of local models from another point of view. In particular, GIMG and LIMG, as a special case of GLIMG, their performance are represented at 0 and 1, respectively. It is obvious that Figure 3(a) and 3(b) share the similar pattern. The NDCG scores first steadily increase and then gradually decline, indicating that only by weighing the global and local models reasonably can the quality of the recommendation be improved.

\begin{table*}[!tp]
	\footnotesize
	\begin{adjustwidth}{-0.95in}{0in}
		\centering
		\caption{Comparison with proposed methods at rank 10 and 50.}
		\begin{tabular}{ccccc|cccc}  \hline
			Dataset    & \multicolumn{4}{c}{ML-1M}                                         & \multicolumn{4}{|c}{Yelp}                                       \\ \hline
			Metric(\%) & NDCG@10        & HR@10          & Precision@10   & Recall@10      & NDCG@10       & HR@10          & Precision@10  & Recall@10     \\ \hline
			GIMG       & 22.27          & 76.26          & 16.84          & 16.55          & 7.62          & 37.21          & 4.51          & 8.51          \\
			LIMG       & 22.34          & 77.50          & 16.44          & 17.37          & 7.35          & 36.48          & 4.42          & 8.28          \\
			GLIMG      & \textbf{23.08} & \textbf{78.42} & \textbf{17.08} & \textbf{17.73} & \textbf{7.81} & \textbf{37.37} & \textbf{4.60} & \textbf{8.76}  \\ \hline
		\end{tabular}
		
		\centering
		\begin{tabular}{ccccc|cccc}  \\ \hline
			Dataset    & \multicolumn{4}{c}{ML-1M}                                        & \multicolumn{4}{|c}{Yelp}                                         \\  \hline
			Metric(\%) & NDCG@50        & HR@50          & Precision@50  & Recall@50      & NDCG@50        & HR@50          & Precision@50  & Recall@50      \\ \hline
			GIMG       & 27.29          & 91.81          & 9.79          & 37.64          & 12.79          & 68.66          & 2.82          & 23.37          \\
			LIMG       & 27.96          & 93.69          & 9.50          & 39.62          & 12.50          & 68.34          & 2.78          & 23.16          \\
			GLIMG      & \textbf{28.66} & \textbf{93.77} & \textbf{9.90} & \textbf{40.17} & \textbf{13.20} & \textbf{70.27} & \textbf{2.89} & \textbf{24.27}  \\ \hline
		\end{tabular}
		
	\end{adjustwidth}
\end{table*}

\begin{figure*}[!t]\vskip -0.05 in
	\begin{center}
		{\subfigure[MovieLens-g]
			{\includegraphics[width=0.45\textwidth]{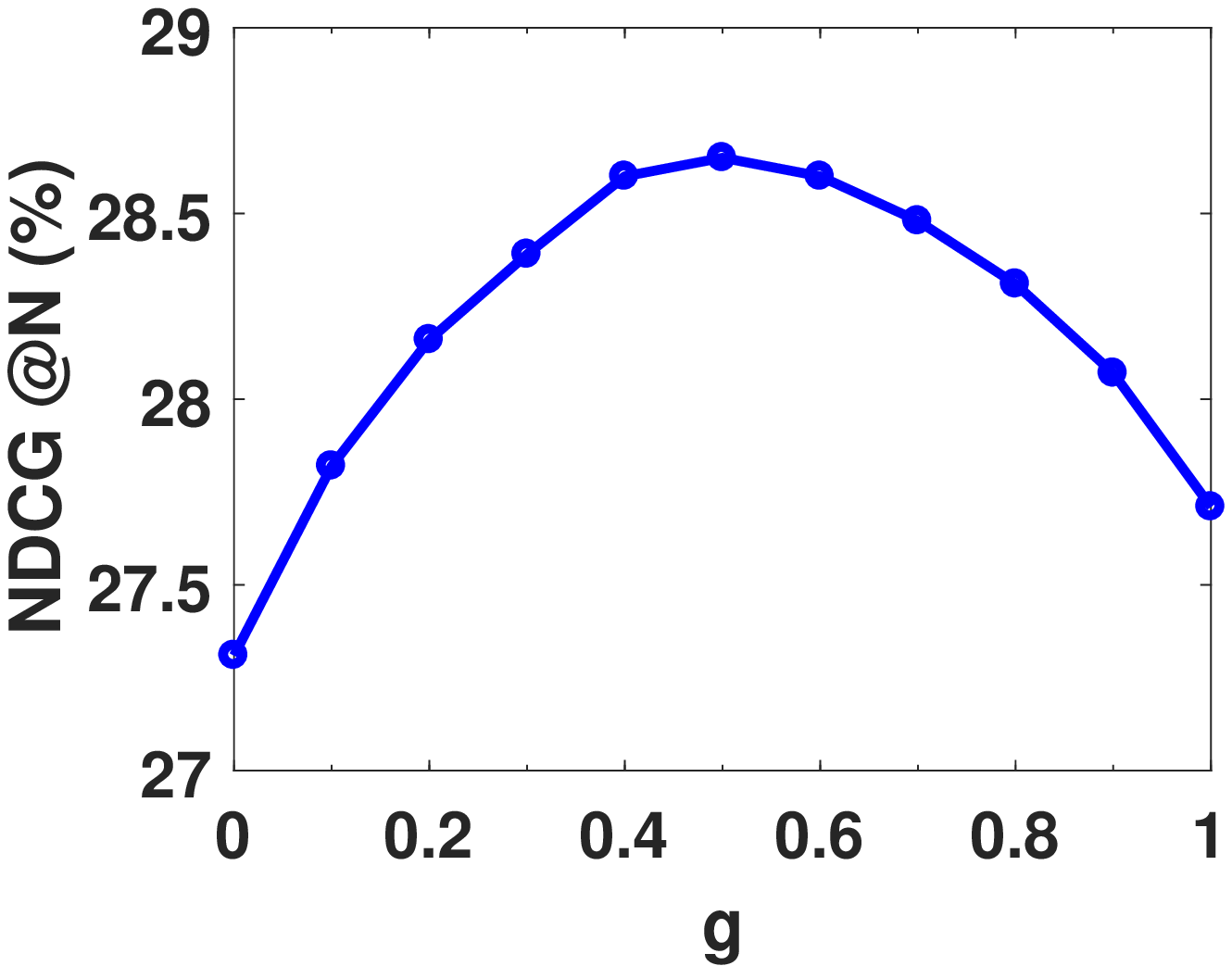}\label{fig4:a}}}
		{\subfigure[Yelp-g]
			{\includegraphics[width=0.45\textwidth]{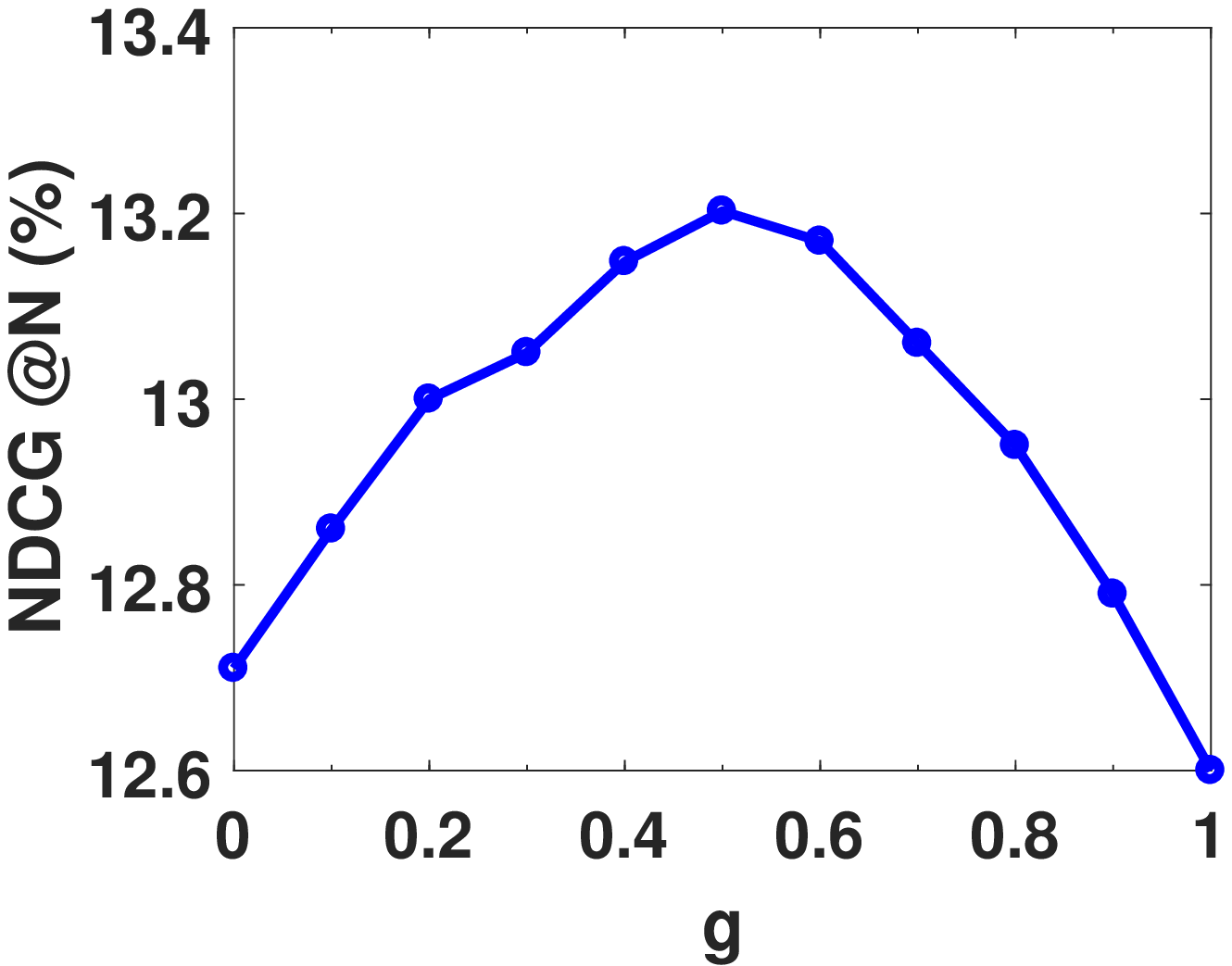}\label{fig4:b}}}
		\vspace*{8pt}
		\caption{The influence of the weight between the global and local model evaluated by NDCG at rank 50. }
		\label{fig4}
	\end{center}\vskip -0.15 in
\end{figure*}

\begin{figure*}[!t]\vskip -0.05 in
	\begin{center}
		{\subfigure[MovieLens-k]
			{\includegraphics[width=0.45\textwidth]{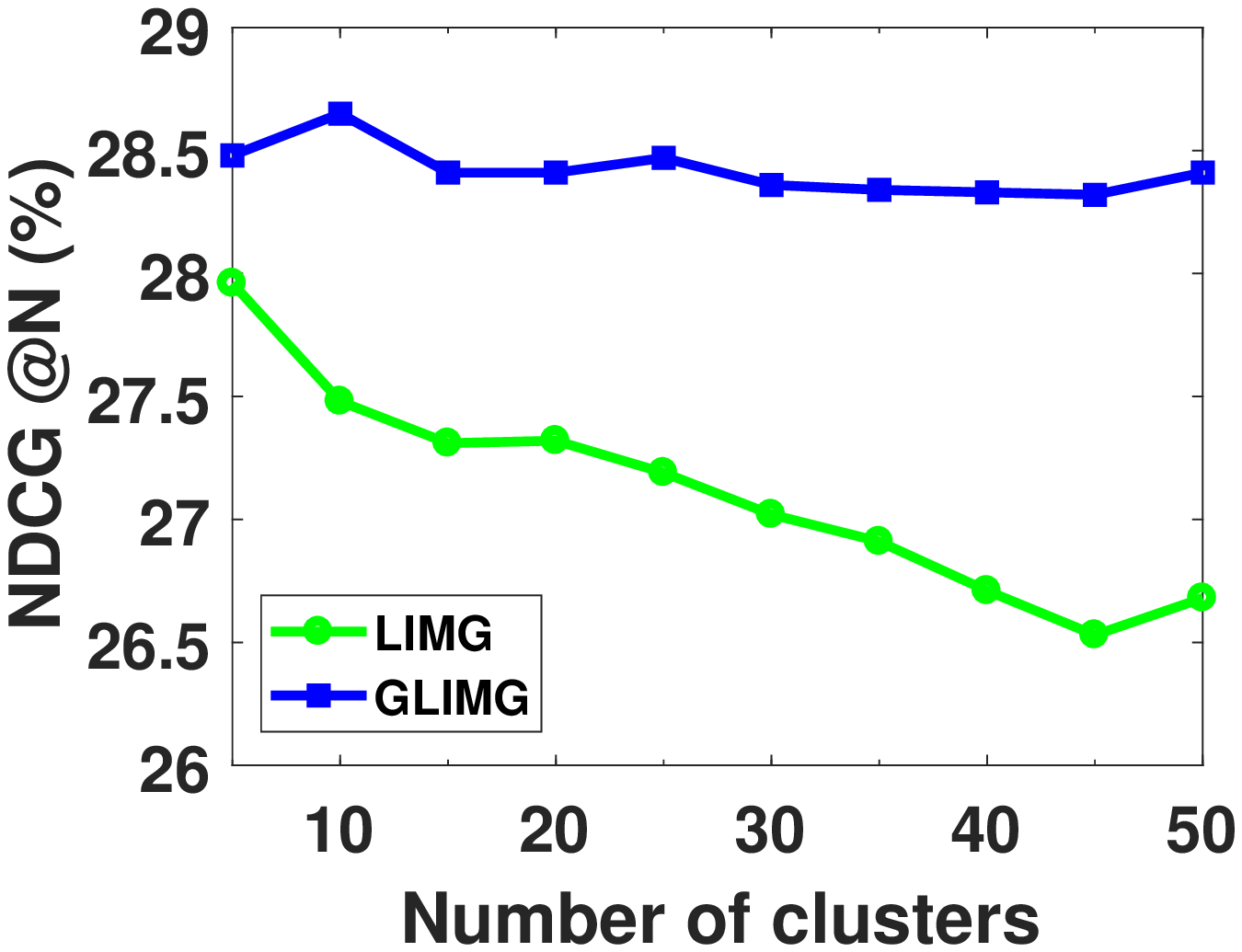}\label{fig5:a}}}
		{\subfigure[Yelp-k]
			{\includegraphics[width=0.45\textwidth]{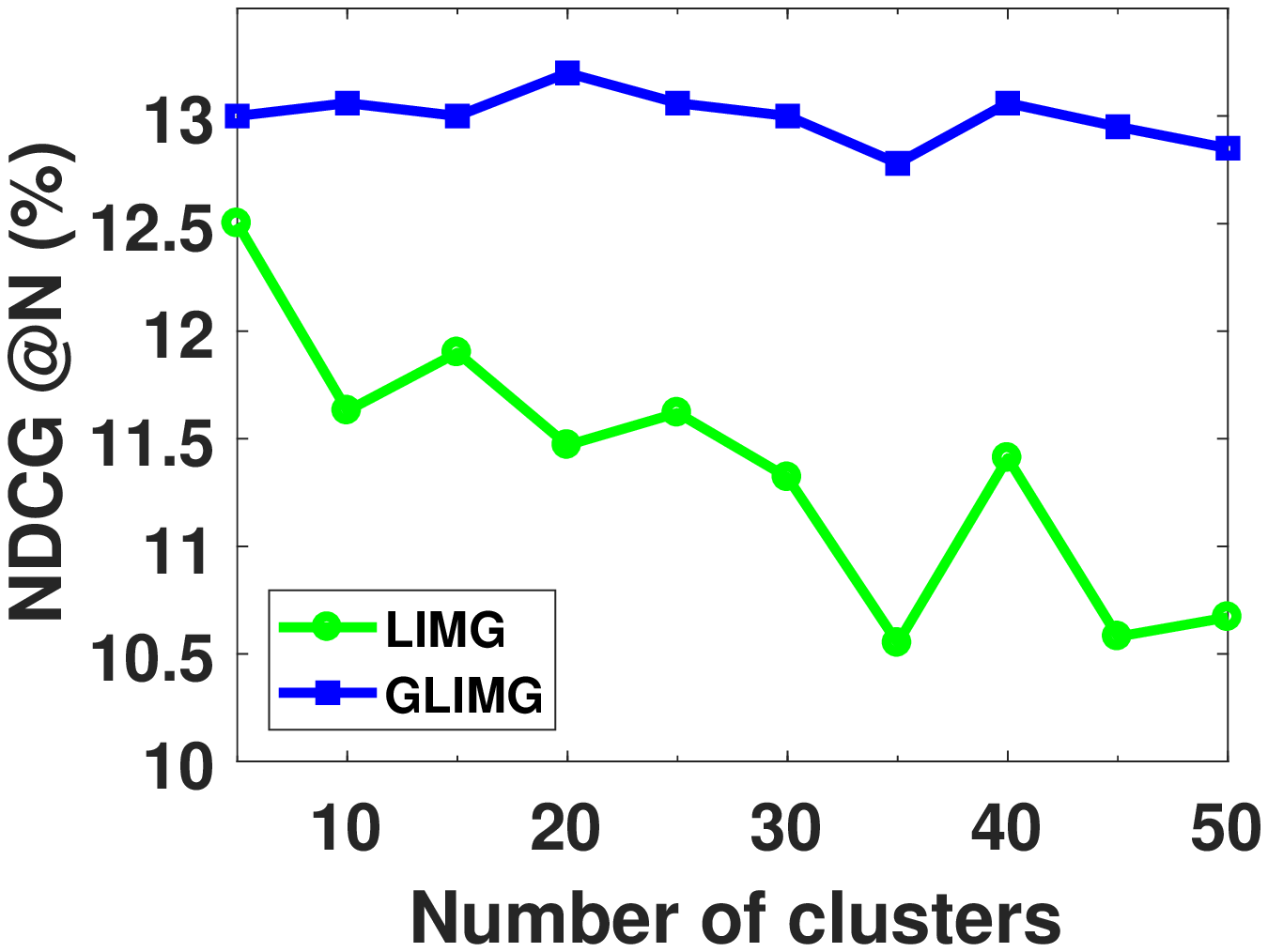}\label{fig5:b}}}
		\vspace*{8pt}
		\caption{The influence of the number of clusters evaluated by NDCG at rank 50.}
		\label{fig5}
	\end{center}\vskip -0.15 in
\end{figure*}

\begin{figure*}[!h]\vskip -0.05 in
	\begin{center}
		{\subfigure[MovieLens-$\sigma$]
			{\includegraphics[width=0.32\linewidth]{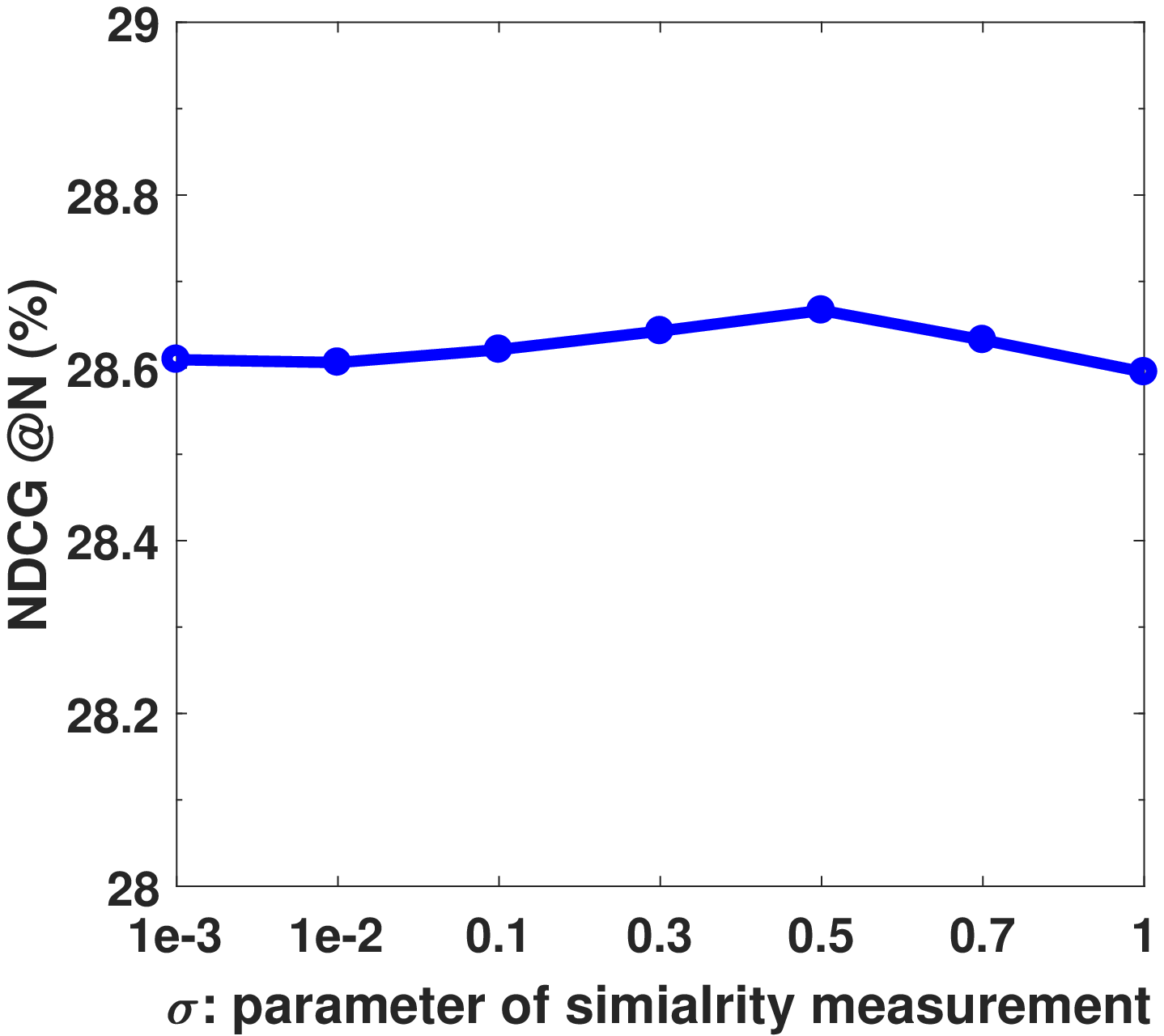}\label{fig6:a}}}
		{\subfigure[Yelp-$\sigma$]
			{\includegraphics[width=0.32\linewidth]{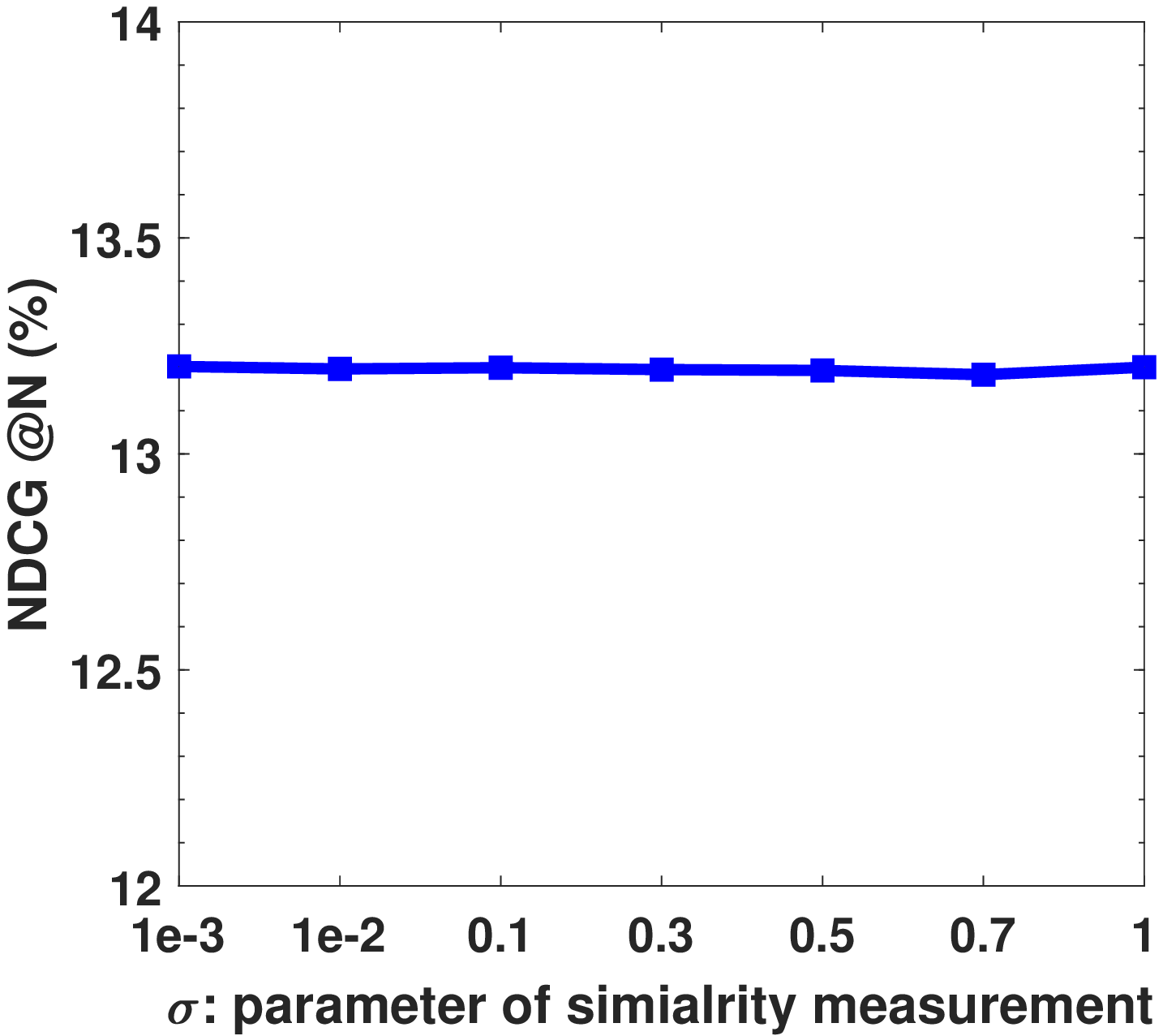}\label{fig6:b}}}
		{\subfigure[MovieLens-$\mu$]
			{\includegraphics[width=0.32\linewidth]{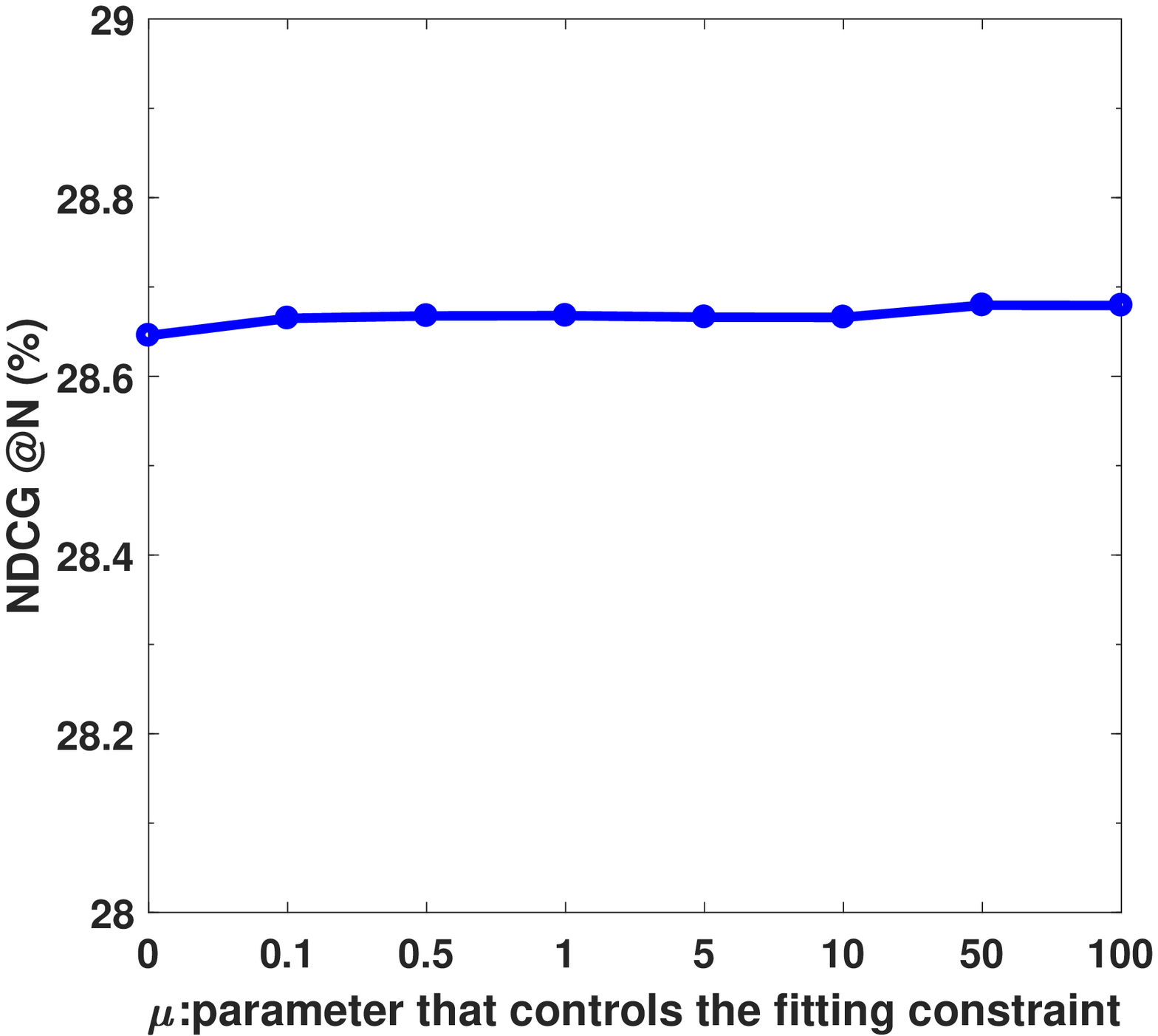}\label{fig6:c}}}
		{\subfigure[Yelp-$\mu$]
			{\includegraphics[width=0.32\linewidth]{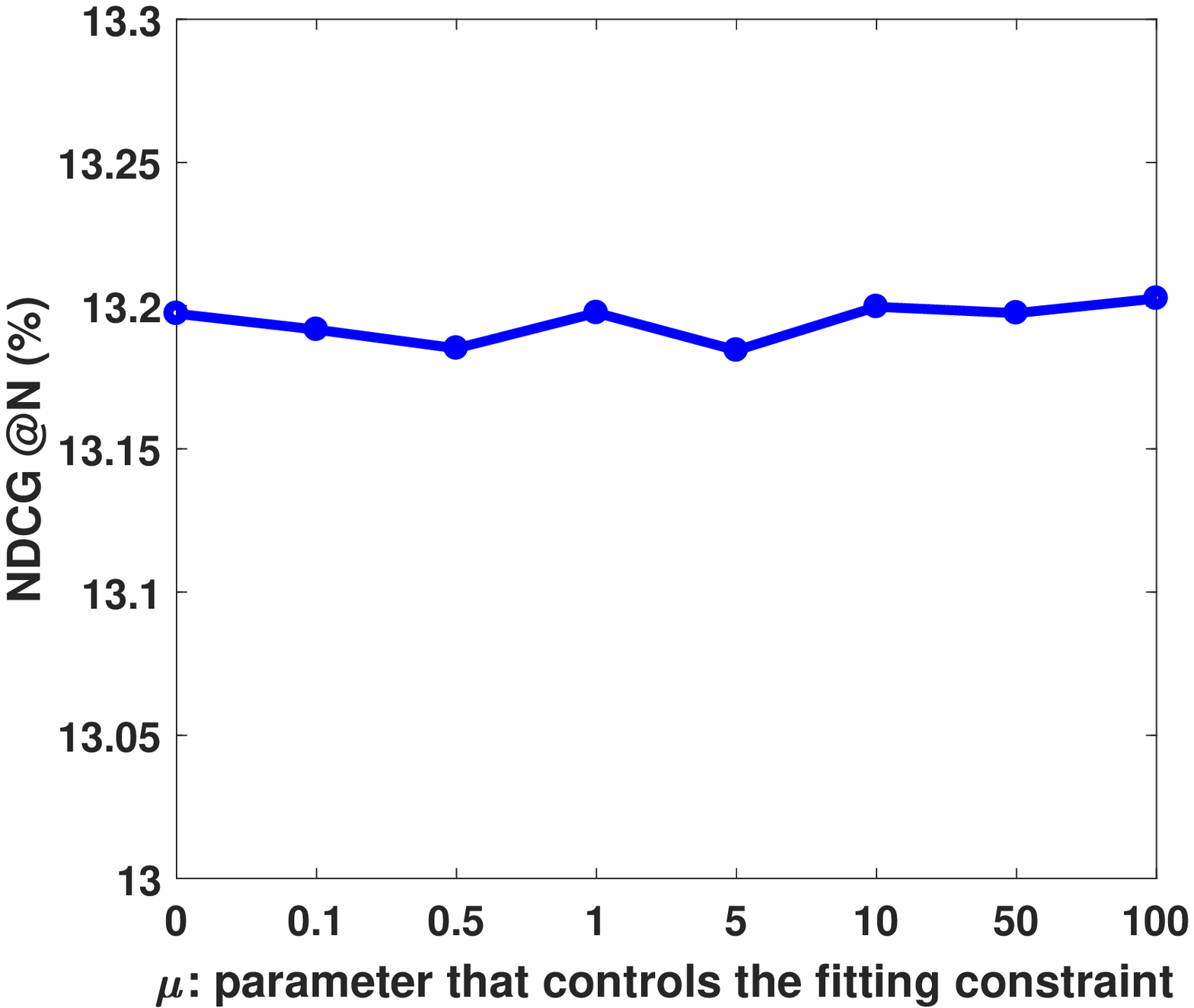}\label{fig6:d}}}
		{\subfigure[MovieLens-$\gamma$]
			{\includegraphics[width=0.32\linewidth]{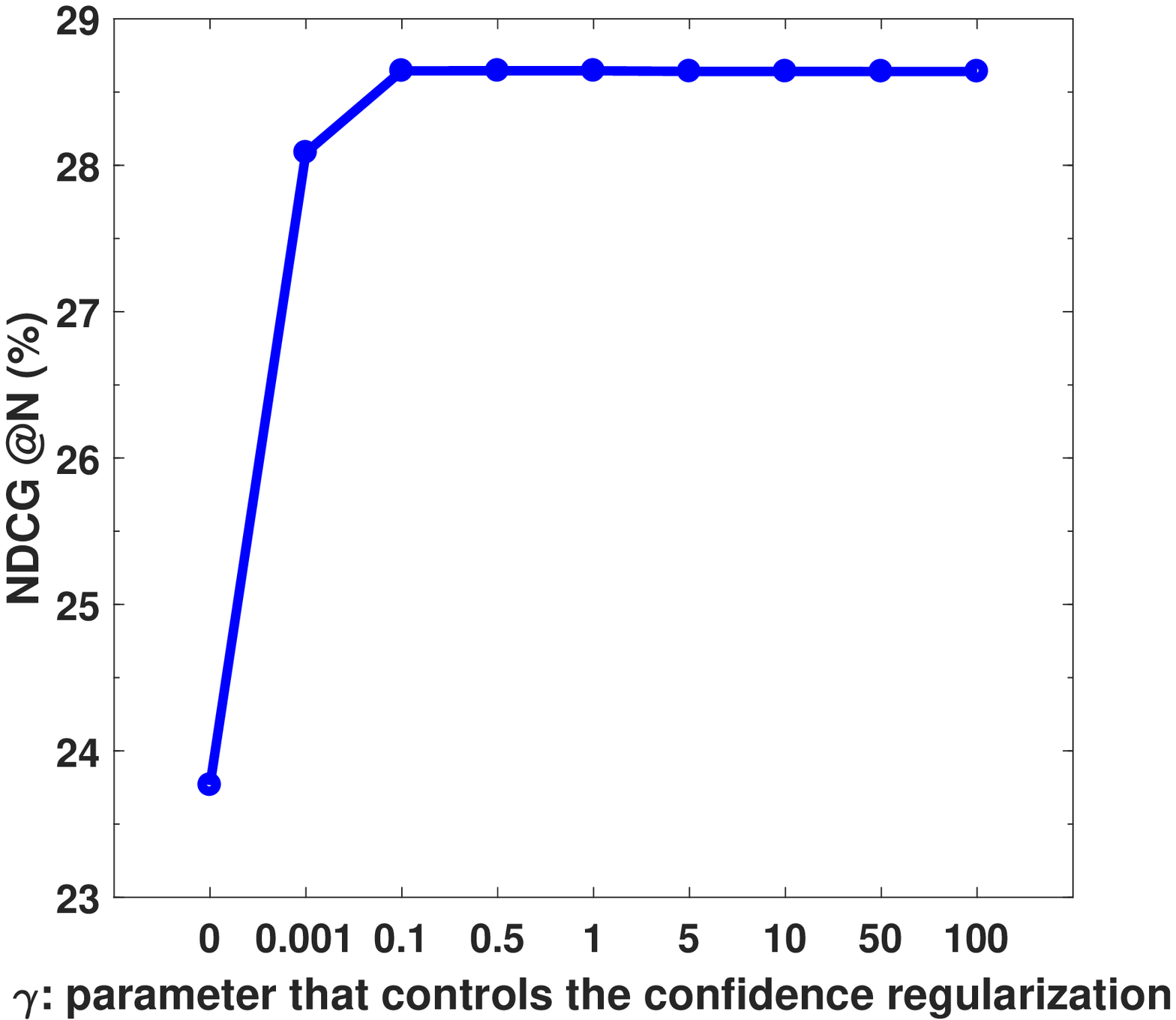}\label{fig6:e}}}
		{\subfigure[Yelp-$\gamma$]
			{\includegraphics[width=0.32\linewidth]{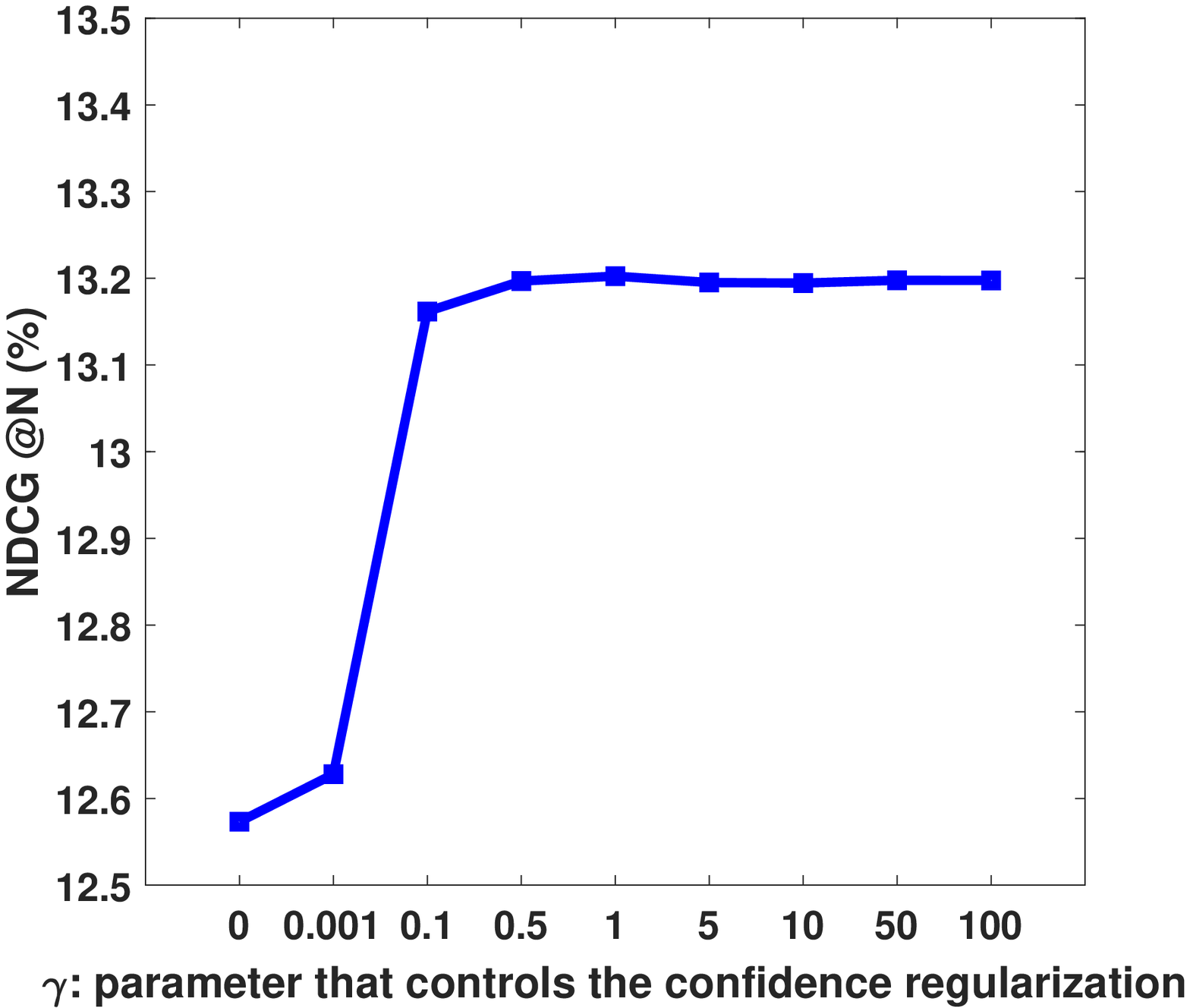}\label{fig6:f}}}\vskip -0.11 in
		\vspace*{8pt}
		\caption{The influence of $\sigma$, $\mu$ and $\gamma$ evaluated by NDCG at rank 50.}
		\label{fig6}
	\end{center}\vskip -0.15 in
\end{figure*}

We also discover the effect of the number of local models for GLIMG and its variant LIMG on both datasets, which is plotted in Figure 4, showing the results from 5 to 50. We can see that LIMG only performs well when the number of clusters is small and with the increase of the number of clusters, the performance evaluated by NDCG decreases in general. It is worth noting that the GLIMG remains very stable with the change of the number of clusters while achieving better performance compared to LIMG. This shows such combination of global item graph and multiple local item graphs will not introduce the instability of local models while improving the quality of recommendation. Another interesting finding is that GLIMG and LIMG are able to achieve impressive performance by constructing few clusters (e.g. 5 clusters), which ensure the computational efficiency of our method.

\subsection{Sensitivity Analysis}

Figure 5 depicts the different values of $\sigma$, $\mu$ and $\gamma$ evaluated by NDCG on dataset MovieLens and Yelp. In this work, $\mu$ and $\gamma$ are chosen from $\lbrace$0,0.1,0.5,1,5,10,50,100$\rbrace$ while $\sigma$ is chosen from $\lbrace$0,0.001,0.01,0.1,0.3,0.5,0.7,1$\rbrace$. In general, the performance of GLIMG is not sensitive to the value of $\sigma$ and $\mu$. We first focus on Figure 5(a) and 5(b) which present the effect of $\sigma$. The performance increases slowly when $\sigma$ is small, and the performance begins to decrease after a certain point on MovieLens dataset. Meanwhile, the performance of $\sigma$ remains stable on the Yelp dataset. Figure 5(c) and 5(d) show the influence of $\mu$ evaluated by NDCG. The results highlight that the performance of $\mu$ is very stable on the MovieLens dataset, while there are slightly fluctuations on the Yelp dataset.

As for $\gamma$, when the $\gamma$ is large enough, it is very stable on both datasets and no significant fluctuations can be observed. It is obvious that when $\gamma$ is small, the performance of GLIMG degrades significantly on both datasets, indicating the third regularization term in Eq. (6) is an essential part for our model to achieve personalized recommendation. The sensitivity test of $\gamma$ shows the effectiveness of our proposed hypothesis 3.

\section{CONCLUSION AND FUTURE WORKS}
In this paper, we propose a novel recommendation approach named GLIMG, which aims to construct multiple local item graphs along with a global graph for improving the performance of graph-based models on the top-N recommendation task. In this way, not only local tastes of different user subgroups but also the global taste shared by all the users could be captured. Extensive experimental results on real-world datasets show that GLIMG consistently outperforms the state-of-the-art counterparts. Meanwhile, we also find that the instability of local models will not be introduced in GLIMG, and a small number of clusters (e.g. 5) is enough under our experimental setting. Since similarity measures are extremely important for graph-based recommendation models, in future work, we will explore if there exist better ways to measure the similarity between items. Another direction is to develop more effective user clustering methods for graph-based recommendation approaches. This enables local graph models to capture more accurate local user preferences for top-N recommendation. In addition, the attempt of applying matrix approximation techniques to estimate the inverse of matrix $(I+\alpha L)$ will be studied to reduce the offline time complexity.

\section{ACKNOWLEDGEMENT}
This work was supported in part by the A*STAR-NTU-SUTD Joint Research Grant RGANS1905, and in part by the Singapore Institute of Manufacturing Technology-Nanyang Technological University (SIMTech-NTU) Joint Laboratory and Collaborative Research Programme on Complex Systems. In addition, we would like to thank anonymous reviews for their valuable suggestions.









\bibliography{mybibfile}
\bibliographystyle{elsarticle-num-names}

\end{document}